\documentclass[11pt, a4paper]{article}
\usepackage{parskip}
\usepackage{latexsym}
\usepackage{times}
\usepackage{color}
\usepackage{graphicx}
\usepackage{amsmath}
\usepackage{setspace}

\date{}
\begin{document}
\title{Advection, diffusion and delivery over a network.}
\author{Luke L.M. Heaton$^{1,2}$, Eduardo L\'{o}pez$^{2,3}$, \\
Philip K. Maini$^{3,4,5}$, Mark D. Fricker$^{3,6}$, Nick S. Jones$^{2,3,5}$}

\maketitle

\begin{center}
{\footnotesize $^{1}$ LSI DTC, Wolfson Building, University of Oxford, Parks Road, Oxford, OX1 3QD, UK \\
$^{2}$ Physics Department, Clarendon Laboratory, University of Oxford, \\ Parks Road, Oxford, OX1 3PU, UK \\
$^{3}$ CABDyN Complexity Centre, Sa\"{\i}d Business School, University of Oxford, \\ Park End Street, Oxford, OX1 1HP, UK \\
$^{4}$ Centre for Mathematical Biology, Mathematical Institute, University of Oxford, \\ 24-29 St Giles', Oxford, OX1 3LB, UK \\
$^{5}$ Oxford Centre for Integrative Systems Biology, Department of Biochemistry, \\ University of Oxford, South Parks Road, Oxford, OX1 3QU, UK \\
$^{6}$ Department of Plant Sciences, University of Oxford, South Parks Road, Oxford, OX1 3RB, UK}
\end{center}

\vspace{5mm}

Many biological, geophysical and technological systems involve the transport of resource over a network. In this paper we present an algorithm for calculating the exact concentration of resource at any point in space or time, given that the resource in the network is lost or delivered out of the network at a given rate, while being subject to advection and diffusion. We consider the implications of advection, diffusion and delivery for simple models of glucose delivery through a vascular network, and conclude that in certain circumstances, increasing the volume of blood and the number of glucose transporters can actually decrease the total rate of glucose delivery. We also consider the case of empirically determined fungal networks, and analyze the distribution of resource that emerges as such networks grow over time. Fungal growth involves the expansion of fluid filled vessels, which necessarily involves the movement of fluid. In three empirically determined fungal networks we found that the minimum currents consistent with the observed growth would effectively transport resource throughout the network over the time-scale of growth. This suggests that in foraging fungi, the active transport mechanisms observed in the growing tips may not be required for long range transport.

\textbf{Keywords: Transport networks; fungal networks; vascular networks; advection-diffusion.}


\section{Introduction}
Many biological, geophysical and technological systems involve the transport of material over a network by advection and diffusion \cite{Avraham, Bunde, Kirkpatrick, Makse, Sahimi}, and it is common that this material can leave, decay, be lost, consumed or delivered as it propagates. Indeed, fluid transport systems are found in the vast majority of multicellular organisms, as the component cells of such organisms require resources for metabolism and growth, and diffusion alone is only an effective means of exchange at microscopic length scales \cite{LaBarbera}.  Molecules of interest are carried by advection and diffusion through the cardio-vascular networks of animals \cite{Beard1, Beard2, Goldman2, Goldman1, Kirkpatrick, Sherman, Shipley, Szczerba, Truskey}, the mycelial networks of fungi \cite{Cairney, Jennings}, the xylem and phloem elements of tracheophytes (vascular plants) \cite{McCulloh, Sack, Thompson}, and various body cavities of many different animals. For example, oxygen is transported through the lungs of mammals and the trachea of insects, while suspension feeding animals (including sponges, clams, brachiopods, many arthropods, fish, ascidians and baleen whales) pass water through various chambers of their bodies, capturing the organic particles that are needed for survival \cite{LaBarbera}. Similar mechanisms of transport are also found in geological and technological systems, such as rivers and drainage networks \cite{Banavar1}, gas pipelines, sewer systems and ventilation systems \cite{Truskey, West}.

In all of these cases the particles of interest diffuse within a moving fluid, which is constrained to flow within a given network. The bulk movement of fluid is referred to as advection, convection or mass flow, and in general the fluid in question travels with a mean velocity that varies over the network. The mean velocity of fluid flow may vary by several orders of magnitude, as, for example, the velocity of human blood drops from $1 \textrm{m s}^{-1}$ in the aorta to around $1 \textrm{mm s}^{-1}$ in the capillaries \cite{Butti, Daley}. Given a network and a distribution of velocities, we may wish to calculate how an initial distribution of resource changes over time. For example, we might want to know how a patch of pollutant will spread within a drainage network \cite{Avraham, Makse, Sahimi}, how a drug will spread within the cardio-vascular system \cite{Beard1, Beard2, Goldman2, Goldman1, Sherman, Shipley, Truskey}, or how nutrients will be translocated within a fungal network \cite{Cairney, Jennings}. In this paper, we consider the particular cases of modelling the delivery of glucose via a vascular network, and modelling the translocation of nutrients in a fungal network. 

Koplik et. al. \cite{Koplik} describe an effective method for calculating the exact moments of the transit times for a neutral tracer across an arbitrary network that contains a flowing medium, but which initially contains no tracer. We have advanced their methods to handle resources that may be consumed or delivered out of the network, while the resource that remains in the network moves by advection and diffusion. More specifically, we suppose that each edge in the network has a local delivery rate $R_{ij}$, which represents the probability per unit time that any given unit of resource will be consumed, lost or delivered out of the network. The effect of including a delivery term can be significant and somewhat counter-intuitive: we will see, for example, that there are circumstances in which increasing the number of blood vessels in a region can actually decrease the amount of glucose that is delivered to that region (Section \ref{idealized vascular networks}). This problem is of particular bio-medical interest, as glucose delivery is essential to the survival of tumours and healthy tissue \cite{Carmeliet1, Chaplain, Kirkpatrick, Shipley, Truskey}. As we shall see, to appreciate how the number of blood vessels in a region effects the total rate of glucose delivery, it is essential that we consider both the rate of delivery of resource out of the network and the topology of the transport network itself. 

To enable the assessment of the transport characteristics of arbitrary networks, with velocities that may vary over several orders of magnitude, we have developed a mathematical methodology that operates in Laplace space. We were initially motivated to develop this algorithm by our interest in fungal networks. Peculiarly, the translocation of resource within fungal networks is much less well studied than transport in the other major multi-cellular kingdoms of life, but the ability of fungal colonies to translocate resources is ecologically critical \cite{Gadd}. The relative roles of mass-flows (advection), diffusion and active transport are very poorly understood. Independent of exclusively fungal questions, fungal systems have the benefit that the network is accessible, and development can be readily followed through a sequence of images. 

We have structured this paper to bring out the applications of our approach. As a consequence, a good part of the mathematical detail is in the  Appendix. Although an important part of the paper's new results and machinery is in the Appendix, familiarity with that material is not needed to understand the results that we discuss in the main text. The outline of this paper is as follows: Preliminary assumptions and the fundamental equations governing advection, diffusion and delivery are discussed (Sections \ref{Preliminary assumptions} and \ref{fundamental equations}), and we stress the importance of the relevant time-scales for advection, diffusion and delivery (Section \ref{time-scales and transport}). We have developed a mixed method that enables us to calculate the exact concentration of resource at any point in space or time in an arbitrary network, and in the  Appendix we describe two efficient algorithms for updating the concentrations in a network over time, given an arbitrary, stepwise constant initial condition, and any number of point sources. In Section \ref{advection etc in Laplace space} we give a brief account of the convenience of solving the fundamental equations in Laplace space, and outline the key ideas and equations of our approach. Alternative methods for solving the fundamental equations are outlined in Section \ref{Alternative methods}.

Finally, we apply our algorithms to a number of test cases, including a model of glucose transport in an idealized vascular network (Section \ref{idealized vascular networks}). We are motivated to understand how the geometry of a vascular network impacts upon the total rate of glucose delivery, as the effective use of anti-angiogenic drugs depends upon understanding the relationship between vascular pruning and nutrient delivery. We also apply our algorithm to a model of resource translocation across empirically determined, growing fungal networks (Section \ref{fungal networks}). We note that changes in fungal volume requires the movement of fluid: for example, the cytoplasm in a growing hyphal tube moves forward with the growing tip \cite{Lew}. In Section \ref{fungal networks} we use our algorithm to investigate whether these growth induced currents are sufficient to supply the tips with the resources they require. In three empirically determined fungal networks we found that the minimum currents consistent with the observed growth would effectively transport resource from the inoculum to the growing tips over the time-scale of growth. This suggests that the active transport mechanisms observed in the growing tips of fungal networks may not be required for long range transport.

\section{Further details} \label{basic principles}
\subsection{Preliminary assumptions} \label{Preliminary assumptions}
We are interested in calculating the distribution of resource across a network of tubes, where the resource in question has a molecular diffusion coefficient $D_{m}$, and where we are given four essential properties for each edge in the network (see Fig. 1). The edge connecting nodes $i$ and $j$ has:
\begin{enumerate}
\item A cross-sectional area, denoted $S_{ij}(t)$. We assume that $S_{ij}(t)$ is piece-wise constant in time, though in  the  Appendix we also consider the more complex case where $S_{ij}(t)$ varies continuously.
\item A length, denoted $l_{ij}$. As the location of the nodes does not vary over time, $l_{ij}$ is constant.
\item A mean velocity, denoted $u_{ij}(t)$. This represents the mean velocity of the fluid in the edge, and we say that $u_{ij}(t)$ is positive if and only if the current flows from node $i$ to node $j$ (so $u_{ij}(t) = -u_{ji}(t)$). By assumption, for each edge $ij$, $u_{ij}(t)$ is piece-wise constant in time.
\item Finally, we suppose that resource in edge $ij$ is delivered out of the network at a rate $R_{ij}$, so if a particle is in $ij$ for a short period of time $\Delta t$, the probability that it is delivered out of the network in that time is $R_{ij} \Delta t$.
\end{enumerate}

\begin{figure}[h!]
\begin{center}
\includegraphics[width=4.5cm]{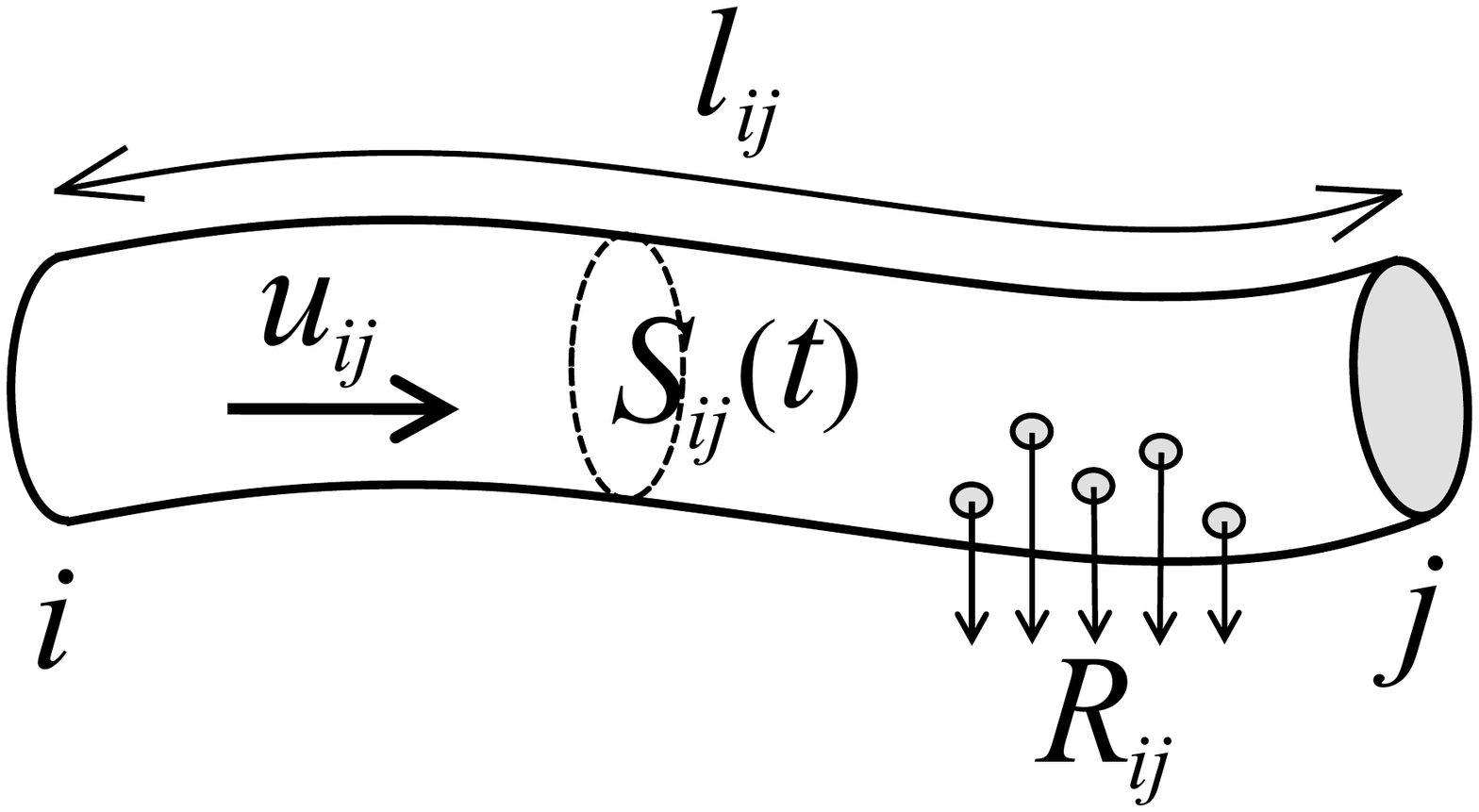}
\caption{\label{single tube}  \textbf{Properties of a single edge in a resource distribution network.}  $S_{ij}(t)$ denotes the cross-sectional area of edge $ij$ at time $t$, $l_{ij}$ denotes the length of the edge, resource and medium flows along the edge with a mean velocity $u_{ij}(t)$, and resource is delivered out of the network at a rate $R_{ij}$. Note that resource travels \emph{along} each edge (and into other edges) by advection and diffusion, but the total rate at which resource in the edge is delivered \emph{out} of the network is simply $R_{ij}$ times the quantity of resource present in the edge. Also note that we do not need to assume that the edges in our network are straight, but we do assume that a single length scale $l_{ij}$ captures the distance that particles must travel to move from $i$ to $j$.}
\end{center}
\end{figure}

While there is a single value for the molecular diffusion coefficient $D_{m}$, the dispersion coefficient $D_{ij}(t)$ may be different for each edge. The value of $D_{ij}(t)$ reflects the tendency of adjacent particles to spread out within $ij$: they not only diffuse along the length of the transport vessels that comprise the edge $ij$, but also diffuse between the slow moving fluid by the edge of the vessels, and the relatively fast moving fluid in the centre of each vessel. 

\begin{figure}[h!]
\begin{center}
\includegraphics[width=7cm]{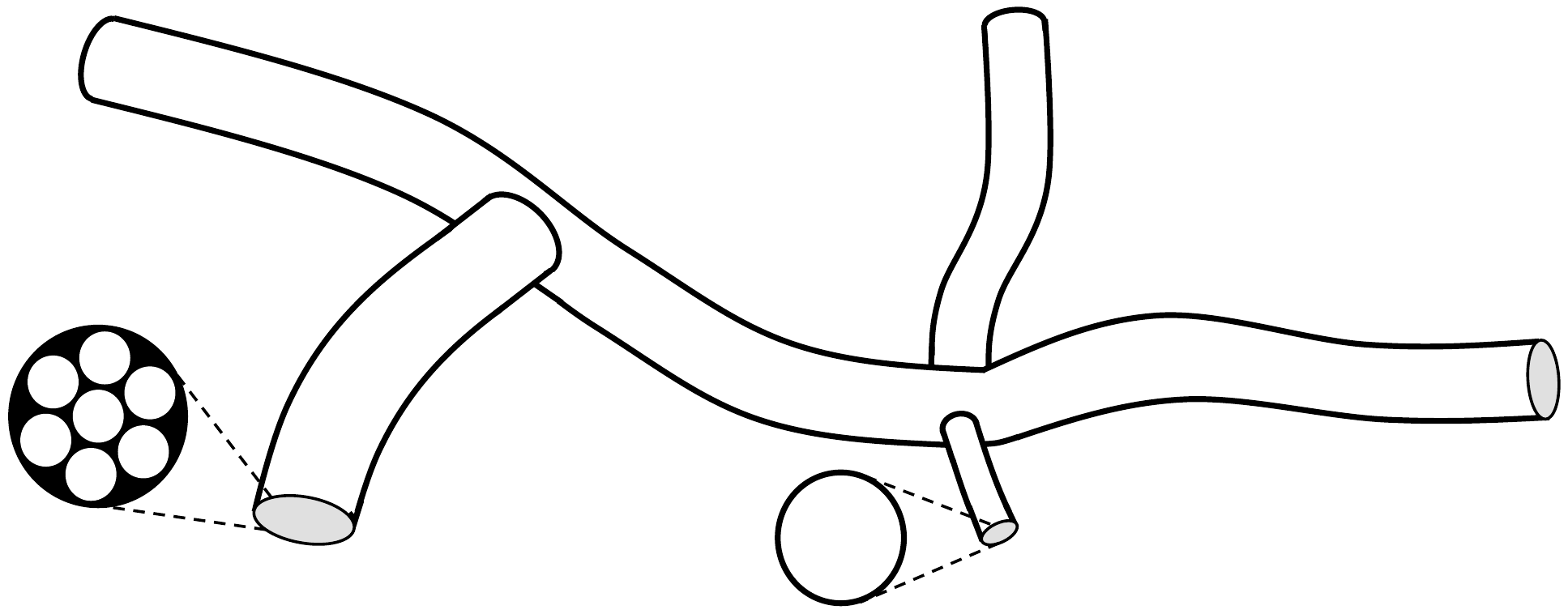}
\caption{\label{network_diagram}  \textbf{Properties of an arbitrary resource distribution network.} Each edge in the network is comprised of a single vessel or a bundle of transport vessels, and each edge has a length $l_{ij}$, a total cross-sectional area $S_{ij}(t)$, a mean velocity of flow $u_{ij}(t)$ and a local delivery rate $R_{ij}$. Each edge also has a dispersion coefficient $D_{ij}(t)$, as described by Equation (\ref{definition_dispersion}). Note that the values of $D_{ij}(t)$ depend on the molecular diffusion coefficient $D_{m}$, the velocities $u_{ij}(t)$ and the radius of the transport vessels within the edge $ij$. The nodes represent the point of contact between the edges: we assume that there is perfect mixing at each node, and we require a consistent concentration at node $i$ whether we consider it to be one end of edge $ij$, or one end of any other edge connected to node $i$.}
\end{center}
\end{figure}

If we consider the case where each edge $ij$ is composed of some number of cylindrical tubes of radius $r_{ij}$ (see Fig. \ref{network_diagram}), and if the Reynold's number is small, we can calculate $D_{ij}(t)$ by using Taylor's dispersion coefficient for laminar flow in a cylindrical tube \cite{Taylor}. This formula tells us that 
\begin{equation}
D_{ij}(t) = D_{m} + u_{ij}(t)^{2} \frac{r_{ij}^{2}}{48D_{m}}.
\label{definition_dispersion}
\end{equation}
In the case of a vascular network $r_{ij}$ is simply the lumen radius of the edge $ij$, so we have $S_{ij} = \pi r_{ij}^{2}$. In plants, fungi or neural tissue each edge in the transport network can be modelled as a bundle of cylindrical tubes; in which case $r_{ij}$ is the characteristic radius of the component transport vessels, and $S_{ij}$ is the total cross-sectional area of the transport vessels.

\subsection{Fundamental equations} \label{fundamental equations}
We suppose that resource is lost or delivered out of the network at a given local rate, while the resource that remains within the network moves by advection and diffusion. Such a process will result in a spatial distribution of resource that changes over time. We only consider longitudinal coordinates along the edge $ij$, using real numbers $x$ to denote distances from node $i$, where $0 \leq x \leq l_{ij}$. Each edge contains a quantity of resource, which must satisfy the one-dimensional advection-diffusion-delivery equation
\begin{equation}
\frac{\partial q_{ij}}{\partial t} + R_{ij} q_{ij} + u_{ij} \frac{\partial q_{ij}}{\partial x} - D_{ij} \frac{\partial^{2} q_{ij}}{\partial x^{2}} = 0,
\label{basic_CDE}
\end{equation}
where $q_{ij}$ is the quantity of resource per unit length, $u_{ij}$ is the mean velocity, $D_{ij}$ is the dispersion coefficient and $R_{ij}$ is the rate at which a unit of resource is lost, or delivered out of the network. In other words, at time $t$ and location $x$, the amount of resource in a $\Delta x$ long slice of the edge is $q_{ij}(x,t) \Delta x$. The distribution of resource within each edge will vary over space and time, but if there is no direct link between the nodes $i$ and $j$, we let $S_{ij}(t) = 0$ and $q_{ij}(x,t) = 0$. This ensures that the sums in the following equations are properly defined for all pairs of nodes $i$ and $j$. 

We wish to find the quantity of resource per unit length $q_{ij}(x,t)$ at a given time $t$. A fundamental assumption underpinning the algorithms described in the  Appendix is that there is perfect mixing at the nodes. In other words, the edge $ij$ is only affected by the rest of the network via the concentrations at nodes $i$ and $j$. This is only a reasonable assumption if the volume of the intersections between edges is negligible in comparison to the volume of the edges themselves. In effect, we assume that the nodes have an infinitesimal volume, so there can be no concentration gradients or boundary layer effects at the junctions between the edges. 

Crucially, the concentration at node $i$ must be consistent across the edges $ij$, $ik$, etc, and we let  $c_{i}(t)$ denote its concentration at time $t$ (amount per unit volume). In other words, for each edge $ij$ we have 
\begin{equation}
c_{i}(t) = \frac{q_{ij}(0,t)}{S_{ij}(t)} \qquad \textrm{and}  \qquad  c_{j}(t) = \frac{q_{ij}(l_{ij},t)}{S_{ij}(t)}, 
\label{definition_ci}
\end{equation} 
where $S_{ij}(t)$ denotes the cross-sectional area at time $t$.

It follows from our assumptions that the concentration profile in edge $ij$ is completely determined by Equation (\ref{basic_CDE}) together with the initial condition $q_{ij}(x,0)$ and the boundary conditions $S_{ij}(t) c_{i}(t)$ and $S_{ij}(t) c_{j}(t)$. By Fick's Law the rate at which resource leaves node $i$ along edge $ij$ is given by
\begin{equation}
J_{ij}(t) =  \bigg[ u_{ij}(t) q_{ij}(x,t) - D_{ij} \frac{\partial q_{ij}(x,t)}{\partial x} \bigg]_{x=0}.
\label{definition_J}
\end{equation}
We assume that resource cannot accumulate at the nodes (as they have zero volume), so any resource that enters node $i$ along edge $ij$ must leave node $i$ along some other edge $ik$. Our framework can accommodate the case where resource is introduced at node $i$ at some given rate $I_{i}(t) > 0$. However, if node $i$ is not an inlet node (that is, a point where resource enters the network), we have $I_{i}(t) = 0$. In either case, Equation (\ref{definition_J}) implies that the net rate at which resource leaves node $i$ is
\begin{equation}
 I_{i}(t) = \sum_{j} \bigg[ u_{ij}(t) q_{ij}(x,t) - D_{ij} \frac{\partial q_{ij}(x,t)}{\partial x} \bigg]_{x=0}.
\label{sums_of_J}
\end{equation}

Equations (\ref{definition_J}) and (\ref{sums_of_J}) describe the current of resource, but we can also consider the current of fluid passing through a given point. Henceforth the term current is reserved for the quantity of resource that passes a given point per unit time, while medium-current refers to the volume of the advecting medium that passes a given point per unit time. The medium-current in edge $ij$ is simply $u_{ij}(t) S_{ij}(t)$, so the net medium-current leaving node $i$ is
\begin{equation}
F_{i}(t) = \sum_{j} u_{ij}(t) S_{ij}(t).
\label{net_current_of_fluid}
\end{equation}

\subsection{Critical time-scales for advection, diffusion and delivery} \label{time-scales and transport} 

For an edge of length $l$, mean velocity $u > 0$, dispersion coefficient $D$ and local delivery rate $R$, there are three critical time-scales: 
\begin{eqnarray*}
t_{A} & = & \frac{l}{u} \textrm{ is the time taken to advect across the edge,} \\ \\
t_{D} & = & \frac{l^{2}}{D} \textrm{ is the mean diffusion time for the edge and} \\ \\
t_{T} & = & \frac{1}{R} \textrm{ is the time-scale of delivery out of the edge.} 
\end{eqnarray*}    
 
 The ratio $\frac{t_{D}}{t_{A}} = \frac{u l}{D}$ is the macroscopic P\'{e}clet number for the edge \cite{Koplik, Truskey}. If $\frac{t_{D}}{t_{A}} \gg 1$ then advection is the dominant form of transport across $ij$, and almost all of the material that leaves $ij$ will pass to locations downstream from $ij$. It is also generally true that in the case of high P\'{e}clet numbers large concentration gradients can persist within each edge \cite{Truskey}. If $\frac{t_{D}}{t_{A}} \ll 1$ then diffusion is the dominant form of transport across $ij$, which means that the concentration within $ij$ will tend to vary smoothly from node $i$ to $j$. 
 
If $t_{T} \ll t_{A}$ and $t_{T} \ll t_{D}$, then the bulk of the resource will be delivered out of the transport network before it transits the edge in question. As a general rule, an efficient transport network will utilize resource over a time-scale $t_{T}$ which is similar to the time-scales over which resource transits the whole network. For example, in the case of vascular networks, the oxygen affinity of haemoglobin varies with body size, and is related to the circulation time for the species in question \cite{Truskey, West}. This makes sense, because if the oxygen affinity of haemoglobin were too low for a given body size, red blood cells would become deoxygenated too rapidly, and too little oxygen would be carried to the tissues distant from the heart and lungs. On the other hand, if a large proportion of the haemoglobin were to remain as oxyhaemoglobin throughout the vascular system, only a small fraction of the oxygen in red blood cells would be transported to the surrounding tissue. As the diffusion coefficient of oxygen is $2 \times 10^{-3} \textrm{ mm}^{2}\textrm{s}^{-1}$ \cite{Truskey} and the velocity of flow in a capillary is about $1 \textrm{mm s}^{-1}$ \cite{Butti}, a capillary of length $1 \textrm{ mm}$ has $t_{A} = 1 \textrm{ s}$ and $t_{D} = 500 \textrm{ s}$. Furthermore, as oxygen is delivered throughout an entire network of capillaries, it follows that $t_{T} \gg t_{A}$.
 

\subsection{Advection, diffusion and delivery in Laplace space} \label{advection etc in Laplace space}

As we explain in the  Appendix, the quantity of resource in each edge is determined by the fundamental Equation (\ref{basic_CDE}) and the concentrations at the nodes. In Laplace space this relationship has a simple algebraic form (Appendix Section 5A).  Furthermore, the Laplace transform of the concentration at the nodes is related to the Laplace transform of the net current passing through each node (Section 5B). We can invert any solutions that we find in Laplace space back into the time domain (Section 5C), and we can also tackle the case of non-zero initial conditions (Section 6). The key idea is that over any time step, the resource in a given edge either reaches one of the nodes at either end of the edge, or it remains within the given edge. Furthermore, the distribution of resource that has not reached either node is equivalent to the distribution of resource that would occur if both nodes were absorbing boundaries. These insights enable us  to formulate an efficient algorithm for calculating  how the spatial distribution of resource changes over time in a fixed, arbitrary network (Section 7). This algorithm couples a network based methodology (Sections 7A and 7B) with analytic solutions for individual edges (Section 7C). By repeated application of this algorithm, we can also find the spatial distribution of resource in a network where the velocities and cross-sectional areas change in a piece-wise constant manner over time (Section 7D). In the Appendix we also consider the more complex case where the cross-sectional areas vary in a continuous manner (Section 8).

The implementation of these algorithms involves the definition of certain constants for each edge (Section 5A). In particular, given any Laplace variable $s$, for each edge $ij$ we let
\begin{equation}
\alpha_{ij}(s) =  \sqrt{u_{ij}^{2} + 4D_{ij}(s + R_{ij})}.
\label{definition_alpha}
\end{equation}
Note that the Laplace variable $s$ represents a rate, and that $\alpha_{ij}(s) = \alpha_{ji}(s)$ is positive, and dimensionally equivalent to speed. Roughly speaking, $\alpha_{ij}(s)$ represents the speed at which resource travels over the time-scale $1/s$, with a correction term to account for delivery.  Since $s$ and $D_{ij}$ are positive and $R_{ij}$ is non-negative, we always find that $\alpha_{ij}(s) > \big| u_{ij} \big|$. The value of $\alpha_{ij}(s)$ depends on $u_{ij}$, $D_{ij}$ and $R_{ij}$ over most time-scales, but for very short time-scales ($s \gg \frac{u_{ij}^{2}}{4 D_{ij}} - R_{ij}$) almost all the movement is due to diffusion,  $\alpha_{ij} \gg u_{ij}$ and $\alpha_{ij}(s) \approx \sqrt{4 D_{ij} (s + R_{ij})}$. 

We let $m$ represent the number of nodes, we let $C_{i}(s)$ represent the Laplace transform of the concentration at node $i$, and $\beta_{ij}(s)$ is a term that reflects the quantity of resource that is initially in edge $ij$, and which leaves $ij$ by passing through node $i$ over the time-scale $1/s$ (Appendix Sections IIA and IIB). In matrix form we have
\begin{equation}
\textbf{M}(s)\bar{C}(s) = \bar{p}(s), \qquad \textrm{where} 
\label{matrix_M_for_CDC}
\end{equation}
\begin{equation*}
\bar{C}(s) = \big\{C_{1}(s), C_{2}(s), \ldots , C_{m}(s) \big\}^{\textrm{T}}, 
\end{equation*}
\begin{equation*}
\bar{p}(s) = \bigg\{ \Upsilon_{1}(s)  + \sum_{j} \beta_{1j}(s), \ldots ,  \Upsilon_{m}(s)  + \sum_{j} \beta_{mj}(s) \bigg\}^{\textrm{T}},   \quad \textrm{and} 
\end{equation*} 
\begin{equation}
\textbf{M}_{ij}(s) =  \Bigg\{	
\begin{array}{cl}
\sum_{k} S_{ik} \bigg[ \frac{u_{ik}}{2} + \frac{\alpha_{ik}(s)}{2  \textrm{ tanh} \big( \frac{ l_{ij} \alpha_{ij}(s)}{2 D_{ij}} \big)} \bigg] & \textrm{ if $i = j$,} \\
& \\
\frac{- S_{ij} \alpha_{ij}(s) e^{\frac{- l_{ij} u_{ij}}{2 D_{ij}} }}{2  \textrm{ sinh}\big( \frac{ l_{ij} \alpha_{ij}(s)}{2 D_{ij}} \big)} & \textrm{ otherwise,}  
\end{array}
\label{defn_M}
\end{equation}

We refer to the matrix $\textbf{M}(s)$ as the propagation matrix, and it contains a row and column for each node in the given network. Given  $\textbf{M}(s)$ and $\bar{p}(s)$ we can calculate $\bar{C}(s)$ using various efficient algorithms, including the stabilized biconjugate gradient method (BiCGStab). In most cases this is the most efficient algorithm to use, as our matrix $\textbf{M}(s)$ is non-symmetric and sparse \cite{Vorst}. Finding $\bar{C}(s)$ is the most time consuming step of our algorithm, but once we have found $\bar{C}(s)$ for $s =  \ln 2/t, 2  \ln 2/t, \ldots, \Omega  \ln 2/t$, we can apply the Gaver-Stehfest algorithm \cite{Abate1, Abate2, Gaver, Stehfest, Zakian1, Zakian2} to find the concentration at time $t$ for every node in the network (further details given in the Appendix).

\subsection{Alternative methods} \label{Alternative methods}
As we outlined in Section \ref{fundamental equations}, there is a system of equations which govern the changing distribution of resource throughout a given network, where the resource in question is subject to advection, diffusion and delivery. There are several methods that could be applied to solve such a system of equations, in addition to algorithms described in the Appendix. We could model the movement of resource by taking a particle based approach, where a large number of particles move across the network, and the path taken by each particle is determined probabilistically, as is the time taken to travel from one node to the next \cite{Sahimi}. 

The problem with such particle based approaches is the challenge of avoiding under-sampling in the regions of the network that contain a low concentration of resource. This problem occurs because, in a finite simulation, the low probability paths are, of course, less well sampled, but the fact that such regions are part of the network may exert a significant effect on the movement of resource, particularly on the higher moments of the transit-times for particles moving across the network \cite{Arcangelis, Koplik, Redner}. Indeed, that is why the dispersion of tracers can be used to probe the structure of networks, and why tracer dispersion plays such a critical role in geophysical surveying techniques \cite{Makse, Groupe, Sahimi}.

Another possible approach is to employ a finite difference scheme. However, in a network where the transport velocities vary over several orders of magnitude, straight forward applications of such an approach are not efficient. The problem is that the time-scale for updating the concentrations is essentially determined by the fastest edge; for stability the distance travelled by advection per time step must be smaller than the spatial resolution (ie. the Courant number must be less than one). Using such a small time step may be very inefficient in the slower moving regions of the network \cite{Courant, Smith}.

\section{Vascular geometry and nutrient delivery} \label{idealized vascular networks}

\subsection{Calculating the total rate of glucose delivery in idealized vascular networks} \label{glucose transport}

We now consider a simple model of glucose moving through a vascular network, where the glucose is `consumed' or transported out of the network by glucose transporters on the surface of the vessels. For the sake of simplicity we assume that the glucose transporters are uniformly distributed over the interior surface of all of the vessels, so the number of transporters per unit length is proportional to the radius of the vessel, and the number of transporters per unit volume of blood is inversely proportional to the radius of the vessel. 

The rate of glucose delivery reflects the frequency of interaction between glucose and the glucose transporters. The kinetics of glucose passing through a transporter is rapid \cite{Marland}, so high concentrations of glucose are required to saturate the transporters. Throughout this section we assume that the glucose concentration is below the carrying capacity ($\textrm{K}_{m}$), and we make the simplifying assumption that the reaction rate is proportional to the concentration of glucose and the concentration of glucose transporters. In other words, we consider the case where the local delivery rate per unit of resource $R_{ij}$ is inversely proportional to the radius of the vessel.

We are interested in the total rate of glucose delivery in different networks of cylindrical tubes, as this quantity corresponds to the total rate at which glucose is transported out of the vasculature and  into the surrounding tissue.  We compare different network geometries by assuming they have one inlet and one outlet node (nodes 1 and 2 respectively). We fix the concentration at node 1, inject some volume of fluid $F$ per unit time at node 1, and remove an equal volume of fluid at node 2. Given the length and radius of each edge, we can calculate the relative conductances, and thereby find the medium-current  flowing through each edge. This enables us to find the velocities $u_{ij}$ as, by definition, the medium-current in each edge is $S_{ij} u_{ij}$. Given the molecular diffusion coefficient for the resource in question, the dispersion coefficients $D_{ij}$ can be found by Equation (\ref{definition_dispersion}).

Numerical simulations indicate that the distribution of resource reaches a steady state. At steady state, the total rate of resource delivery must equal the current of resource entering the network minus the current of resource leaving the network. Furthermore, the fundamental advection, diffusion, delivery Equation (\ref{basic_CDE}) tells us that at steady-state,
\begin{equation}
R_{ij} q_{ij} + u_{ij} \frac{\partial q_{ij}}{\partial x} - D_{ij} \frac{\partial^{2} q_{ij}}{\partial x^{2}} = 0.
\label{fundamental_steady_state}
\end{equation}
It follows that for each edge there must be a pair of constants $A$ and $B$ such that
\begin{equation}
q_{ij}(x)= Ae^{\frac{u_{ij} + \hat{\alpha}_{ij}}{2D_{ij}} x} + Be^{\frac{u_{ij} - \hat{\alpha}_{ij}}{2D_{ij}} x},
\label{steady_state_in_A_and_B}
\end{equation}
\begin{equation}
\textrm{where} \quad \hat{\alpha}_{ij} = \sqrt{u_{ij}^{2} + 4 D_{ij} R_{ij} }.
\label{defn_alpha_0}
\end{equation}

Whatever current of resource and medium we introduce and remove from the given network, the steady state distribution of resource must satisfy Equation (\ref{fundamental_steady_state}). For the sake of simplicity we ignore the process of vascular adaptation whereby vessels dilate, contract or become apoptotic in response to fluid flow and the associated shear wall stress \cite{Alarcon, Owen, Pries, Jain2}, but as our algorithm(s) can be applied to networks with varying cross-sectional areas, we note that such effects could be incorporated into a more complex model. 

Our aim is to compare the efficiency of resource delivery for a range of different networks, and we do this by calculating the total rate of resource delivery for a representative steady state flow of medium and resource. To find such a representative distribution of resource, we suppose that the concentration at node 1 is a fixed constant $k$, and that resource leaves the network by flowing from node 2 into a dummy edge $2n$ (see Fig. \ref{short cut and dead end}). If we suppose that the concentration at node 2 is $c_{2}$ while the concentration at node $n$ is 0, Equation (\ref{steady_state_in_A_and_B}) implies that
\begin{equation*}
A = \frac{- S_{2n} c_{2} e^{\frac{- \hat{\alpha}_{2n}}{D_{2n}} l} }{1 - e^{\frac{- \hat{\alpha}_{2n}}{D_{2n}} l} } \quad \textrm{ and} \quad
B = \frac{ S_{2n} c_{2} }{1 - e^{\frac{- \hat{\alpha}_{2n}}{D_{2n}}l} },
\end{equation*}
where $l$ is the length of the dummy edge $2n$. Letting $l \rightarrow \infty$, we have
\begin{equation}
q_{2n}(x)= S_{2n} c_{2} e^{\frac{u_{2n} - \hat{\alpha}_{2n}}{2D_{2n}} x}.
\label{concentration in dummy edge}
\end{equation}

In this case the flux of resource flowing out of the network at node 2 is 
\begin{equation*}
J_{2n}(x) = \bigg[ u_{2n} q_{2n}(x) - D_{2n} \frac{d}{dx}q_{2n}(x) \bigg]_{x = 0} = F' c_{2},
\end{equation*}
where Equations (\ref{definition_dispersion}), (\ref{defn_alpha_0}) and (\ref{concentration in dummy edge}) tell us that 
\begin{eqnarray}
F' & = & S_{2n} \frac{u_{2n} + \hat{\alpha}_{2n} }{2} \nonumber \\
& = & \frac{F}{2} \bigg(1 + \sqrt{1 + \frac{4 R_{2n} D_{m} S_{2n}^{2} }{F^{2}} + \frac{S_{2n} }{48 \pi D_{m} } } \bigg).
\label{definition of F'}
\end{eqnarray}

Given any network of cylindrical tubes with a specified inlet node 1 and outlet node 2 (see Fig. \ref{short cut and dead end}), and given a molecular diffusion coefficient $D_{m}$ and a local delivery rate $R_{ij}$ for each edge, we can find a spatial distribution of resource that reflects the network's efficiency as a transport system, and we can calculate the total rate of resource delivery in the given case. In particular, it is instructive to calculate the total delivery rate at steady state (denoted $C_{\textrm{tot}}$), which is equal to the total current flowing into the network minus the total current flowing out of the network. We note that 
\begin{equation}
C_{\textrm{tot}} = I_{1}(t) + I_{2}(t) \quad \textrm{for very large $t$},
\label{defn C_tot}
\end{equation} 
and we make a fair comparison between different networks by considering the following:
\begin{enumerate}
\item In each case, we assume that $F_{1}(t) = F$. In other words, at node 1 we inject a volume $F$ of fluid per unit time.
\item We remove an equal volume of fluid from a node 2, so $F_{2}(t) = -F$. 
\item We assume that the flow of fluid is laminar, so the Hagen-Poiseuille equation holds, and the conductance of each edge is proportional to $\frac{S_{ij}^{2}}{l_{ij}}$. 
\item Given the relative conductances of each edge, and given that a medium-current $F$ enters the network at node 1 and leaves the network at node 2, we can calculate the velocites $u_{ij}$ \cite{Heaton, Grimmett}.
\item All the edges are assumed to be cylindrical and composed of a single vessel. As we are given the cross-sectional areas $S_{ij}$ we effectively know the radius of each edge, as well as $u_{ij}$ and $D_{m}$, so we can find the dispersion coefficients $D_{ij}$ by plugging these values into Equation (\ref{definition_dispersion}).
\item  We suppose that the concentration at node 1 is a fixed constant $k$ at all times. This implies that $C_{1}(s) = k/s$.
\item For each edge $ij$, including the dummy edge, we suppose that the delivery rate per unit of resource $R_{ij}$ is inversely proportional to the radius of the vessel. This reflects the assumption that in each vessel there is a fixed density of glucose transporters per unit of surface area.
\item We suppose that the current of resource leaving the network at node 2 is completely determined by the concentration at node 2. More specifically, we let $I_{2}(t) = - F' c_{2}(t)$, where $F'$ is given by Equation (\ref{definition of F'}). Note that the value of $F'$ depends on the same cross-sectional area of the dummy edge, and  in the following section we assume that in each case, $S_{2n} = S_{12}$ (see Fig. \ref{short cut and dead end}).
\item For the sake of simplicity we assume that each network is initially empty, and we calculate the concentrations and total delivery rate for a time point $t$ that is sufficiently large for the system to have reached steady state.
\end{enumerate}

\subsection{Analytic solutions to the total rate of glucose delivery in simple vascular networks}

\begin{figure}[h!]
\begin{center}
\includegraphics[width=8.3cm]{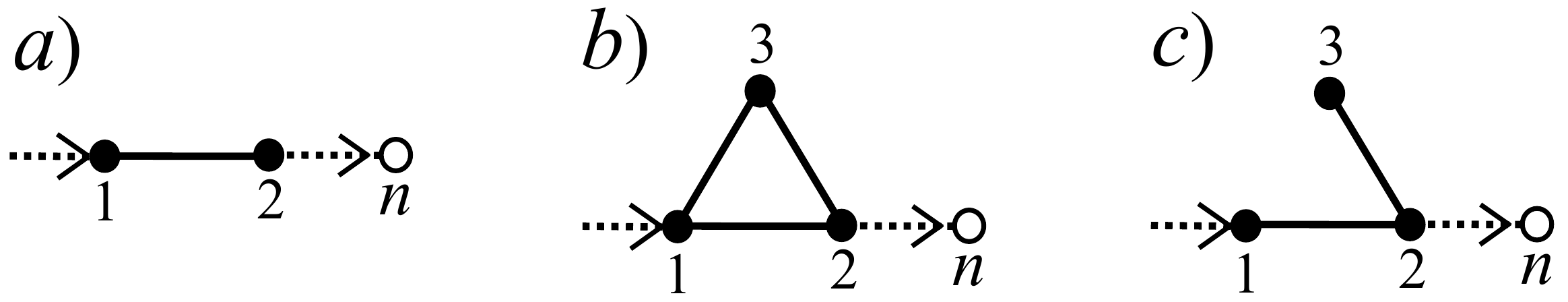}
\caption{\label{short cut and dead end}  \textbf{Network structure has a critical influence on the total rate of resource delivery.} Adding additional vessels may or may not increase the total rate of resource delivery, depending on the extent to which the additional edges change the time taken to transit the network. We illustrate this effect by considering the total rate of delivery in three simple networks, where a current of medium with a fixed concentration of resource is introduced at node 1, and medium and resource leaves the network at node 2 by flowing into the dummy edge $2n$.}
\end{center}
\end{figure}

We begin by considering glucose delivery by a single vessel (see Fig. \ref{short cut and dead end}a), where by definition $F = F_{1}(t) = S_{12} u_{12}$. As we are assuming that $C_{1}(s) = k/s$ and the network is initially empty, $\beta_{ij}(s) = 0$ for every edge $ij$ (see AII), and Equation (\ref{defn_M})  tells us that
\begin{equation*}
\mathbf{M}(s) \bigg(
\begin{array}{c}
k/s \\
C_{2}(s)
\end{array} \bigg)
= \bigg(
\begin{array}{c}
\Upsilon_{1}(s) \\
-F' C_{2}(s)
\end{array} \bigg).
\end{equation*} 
It follows that
\begin{equation}
C_{2}(s) = \frac{- \textbf{M}_{21}(s) k}{s \big(  \textbf{M}_{22}(s) + F' \big)},
\label{C2_single_edge}
\end{equation} 
so we have
\begin{equation}
\Upsilon_{1}(s) = \frac{k}{s} \bigg( \textbf{M}_{11}(s) - \frac{ \textbf{M}_{12}(s) \textbf{M}_{21}(s) }{  \textbf{M}_{22}(s) + F'} \bigg).
\label{Upsilon_single_edge}
\end{equation} 

The approximation $\alpha_{ij}(s) \approx \hat{\alpha}_{ij} = \sqrt{u_{ij}^{2} + 4 D_{ij} R_{ij} }$ is arbitrarily accurate for sufficiently small $s$, so for very small $s$ we can substitute $\textbf{M}_{ij}(0)$ for $\textbf{M}_{ij}(s)$. Hence Equation (\ref{C2_single_edge}) tells us that $C_{2}(s) \propto 1/s$ for very small $s$, and Equation (\ref{Upsilon_single_edge}) tells us that $\Upsilon_{1}(s) \propto 1/s$ for very small $s$. It follows that for sufficiently large $t$,
\begin{eqnarray}
C_{\textrm{tot}} & = & I_{1}(t) + I_{2}(t) = I_{1}(t) - F' c_{2}(t) \nonumber \\
&& \nonumber \\
& = & k \bigg( \textbf{M}_{11}(0) + \frac{ \textbf{M}_{21}(0) F' - \textbf{M}_{21}(0) \textbf{M}_{12}(0) }{  \textbf{M}_{22}(0) + F'} \bigg). \quad
\label{C_tot_in_single_edge}
\end{eqnarray} 

\begin{figure}[h!]
\begin{center}
\includegraphics[width=7.5cm]{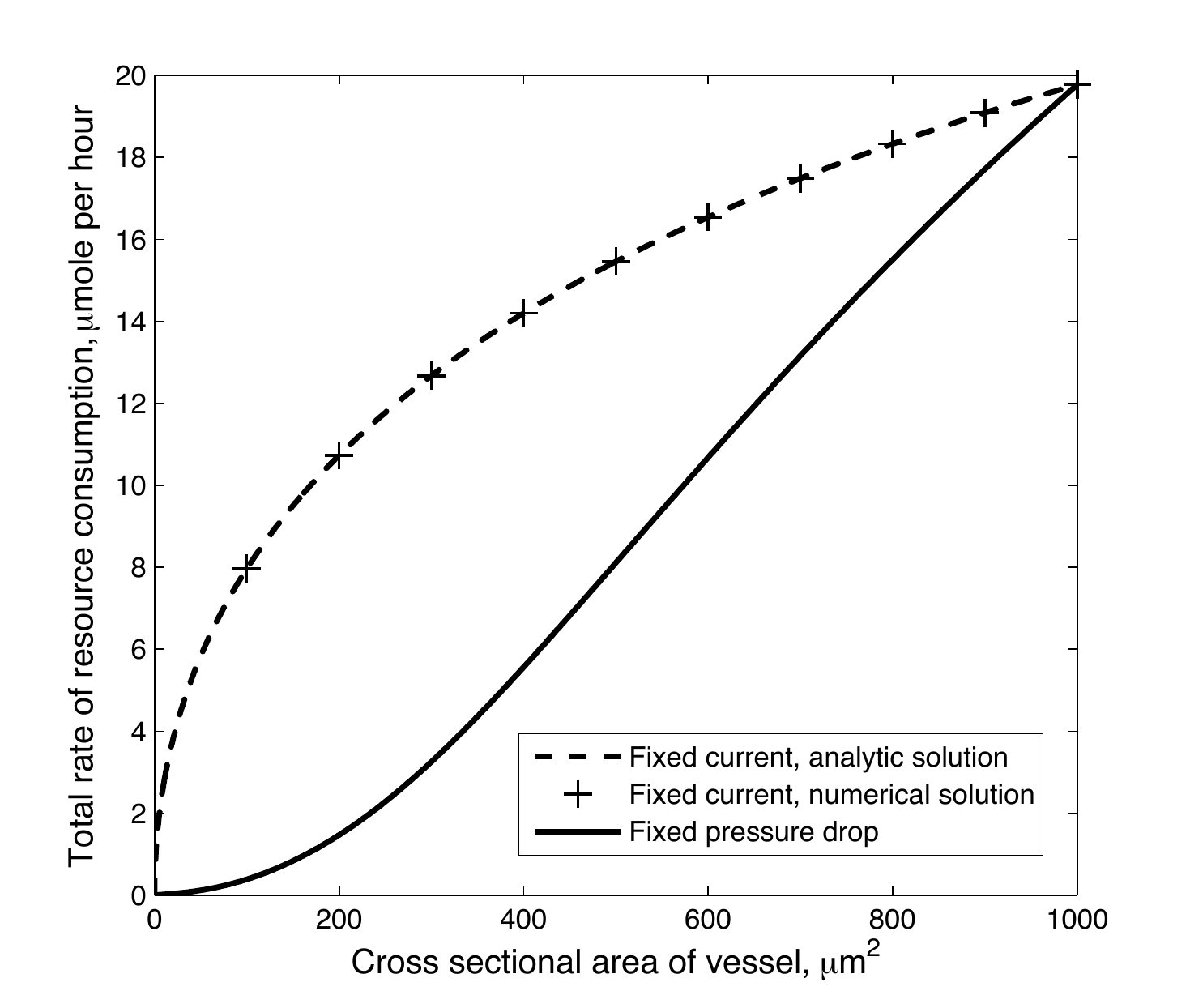}
\caption{\label{single edge delivery}  \textbf{Total rate of resource delivery in a single vessel.} In the case where the medium-current is fixed, the velocity of flow is inversely proportional to the vessel's cross-sectional area $S$. Where the pressure drop is fixed, the velocity of flow is proportional to $S$. Our parameters are such that the velocity is lower in the case where the pressure drop is fixed, up until the point where $S = 1000 \mu\textrm{m}^{2}$, where in either case the velocity $u = 2 \textrm{mm s}^{-1}$, the time-scale of advection is $0.5 \textrm{ s}$ and the time-scale of delivery is $0.63 \textrm{ s}$. Since our parameters imply that the medium velocity is smaller in the case of a fixed pressure drop, we also have a lower mean concentration and a lower total rate of resource delivery. Note that the numerical solution was generated by sampling six points in Laplace space, and applying the Gaver-Stehfest algorithm.}
\end{center}
\end{figure}

Note that the terms $F'$, $u$ and $\hat{\alpha}$ may vary with the cross-sectional area $S$: the  relationship between  $C_{\textrm{tot}}$ and the diameter of vessels is plotted in Fig. \ref{single edge delivery}. To replicate the scales of interest in actual vascular networks, we consider edges of length $l_{12} = 1 \textrm{mm}$, and we let $D_{m} = 6.7 \times 10^{-4} \textrm{mm}^{2}\textrm{s}^{-1}$ (the molecular diffusion coefficient of glucose in water at body temperature). We also let $k = 5 \textrm{mmole/litre}$: a typical value for the concentration of glucose in blood. Finally, we assume that there is a fixed number of glucose transporters  per unit area, so the local delivery rate $R_{12} = \frac{\rho}{\sqrt{S_{12}}} = \frac{0.05}{\sqrt{S_{12}}}$. The numerical value of the parameter $\rho$ reflects the density of transporters, and their affinity for glucose. To illustrate the biologically relevant case, we set $\rho = 0.05 \textrm{mm s}^{-1}$. By choosing this value of $\rho$ we ensure that the concentration of glucose drops significantly between the inlet and the outlet nodes, but in the networks we consider the concentration does not drop by more than an order of magnitude.

To produce the continuous curve shown in Fig. \ref{single edge delivery} we suppose that the medium-current $F = 0.002 \textrm{mm}^{3}\textrm{s}^{-1}$ regardless of the cross-sectional area $S_{12}$. In other words, we suppose that $u_{12} = \frac{F}{S_{12}}$, and find the dispersion coefficient $D_{12}$ by applying Equation (\ref{definition_dispersion}). We have also plotted the case where the pressure drop between nodes 1 and 2 is held constant, rather than the medium-current $F$ (see the dashed curve in Fig. \ref{single edge delivery}). The Hagen-Poiseuille equation states that in the case of laminar flow, conductance should scale with the square of a vessel's cross-sectional area. In other words, in the case of a single edge, maintaining a constant pressure drop is equivalent to setting $u_{12} \propto S_{12}$. Consequently, when the pressure drop is held constant and the cross-sectional area is very low, very little resource enters the network and the total delivery rate is very low. In the case of fixed current, the total delivery rate also drops to zero as the cross-sectional area drops to zero, but in that case it is because the mean velocity is inversely proportional to the cross-sectional area, so when the cross-sectional area is very small, the velocity of flow is very large and a relatively large fraction of the resource leaves the vessel without being delivered out of the transport network. 

\begin{figure}[h!]
\begin{center}
\includegraphics[width=7.5cm]{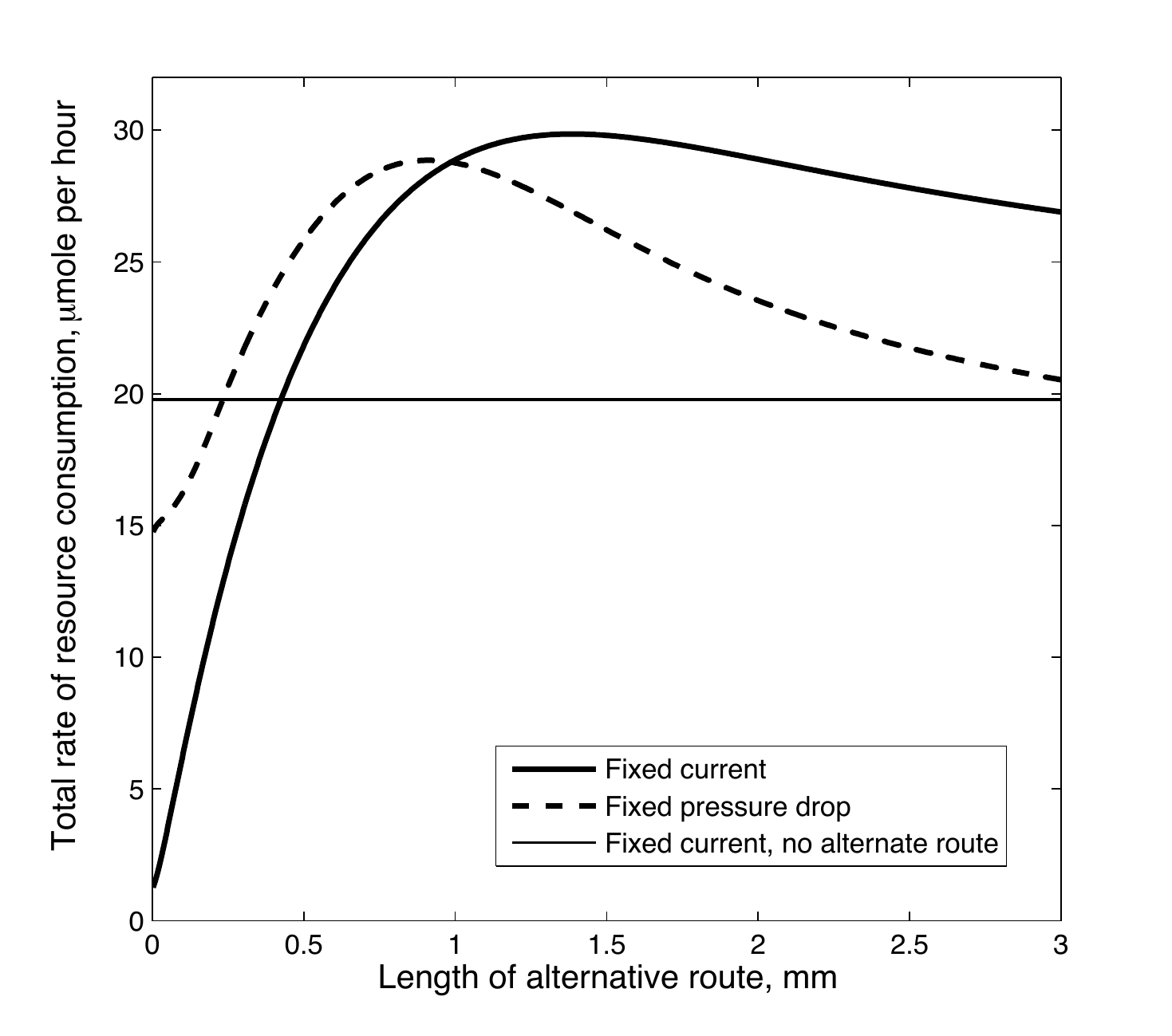}
\caption{\label{delivery_alternate_route} \textbf{Total rate of resource delivery in a network with alternate routes.} The medium-current passing through each route will be proportional to its conductance (see Fig. \ref{short cut and dead end}b). If there is a very short alternative route, its conductance will be very small, the mean transit time will be very small, and so that the total rate of resource delivery will also be small. If the alternative route is sufficiently long, most of the resource entering the alternate route is consumed. Further increases in the length of the alternate route will decreases the total rate of resource delivery, as the medium-current will decrease and so too will the current of resource entering the alternate route. It follows that for a fixed current or a fixed pressure drop, the total rate of resource delivery is at a maximum for some intermediate length of alternate route.}
\end{center}
\end{figure}

We now consider the other networks illustrated in Fig. \ref{short cut and dead end}. By assumption $C_{1}(s) = k/s$, $\Upsilon_{2}(s) = - F' C_{2}(s)$ and $\Upsilon_{3}(s) = 0$. Equation (\ref{defn_M}) relates these terms to the unknowns $\Upsilon_{1}(s)$, $C_{2}(s)$ and $C_{3}(s)$. As in the previous example, this relationship enables us to calculate $C_{\textrm{tot}} = I_{1}(t) - F' c_{2}(t)$. In the case of a triangular network we let $S_{12} = S_{13} = S_{32} = 1000  \mu\textrm{m}^{2}$ and $l_{12} = 1 \textrm{mm}$ (see Fig. \ref{short cut and dead end}b). To illustrate the effect of short-cuts we vary the length of $l_{13} = l_{32}$, and see how it effects $C_{\textrm{tot}}$ (see Fig. \ref{delivery_alternate_route}). As before, $D_{ij}$ is determined by Equation (\ref{definition_dispersion}), $k = 5 \textrm{mmole/litre}$ and we set $R_{ij} = \frac{0.05}{\sqrt{S_{ij}}}$. The velocities $u_{ij}$ are calculated in two different ways: we either fix the total medium-current through the network, or we fix the pressure drop between nodes 1 and 2. In either case, the total rate of resource delivery is at a maximum when the alternate route is of intermediate length (see Fig. \ref{delivery_alternate_route}). 

\begin{figure}[h!]
\begin{center}
\includegraphics[width=7.5cm]{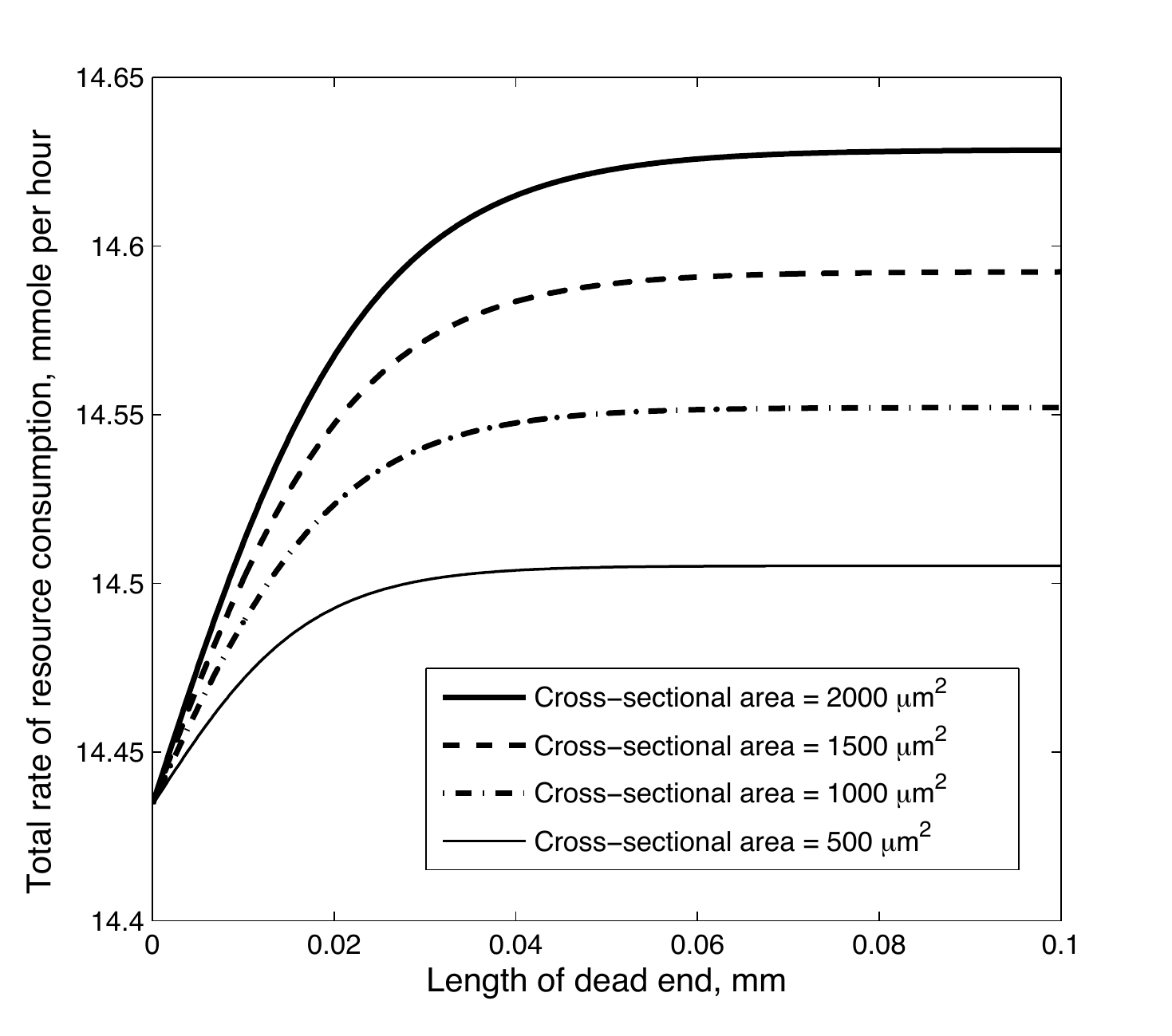}
\caption{\label{delivery_dead_end} \textbf{Total rate of resource delivery in a network with a dead-end.} If the dead-end region is short (see Fig. \ref{short cut and dead end}c), its presence  increases the total rate of resource delivery by an amount that is proportional to both the volume of the dead-end region and the delivery rate per unit of resource $R_{23}$. As we assume that $R_{23} \propto S_{23}^{-\frac{1}{2}}$, it follows that for sufficiently short dead-end regions the increase in the  total rate of resource delivery is proportional to $l_{23} \sqrt{S_{23}}$. The total rate of resource delivery reaches a maximum when the time taken to diffuse the length of the dead-end region is much greater than the time-scale of delivery.}
\end{center}
\end{figure}

As a final example we find $C_{\textrm{tot}} = I_{1}(t) - F' c_{2}(t)$ in the case where our network contains a dead-end (see Fig. \ref{short cut and dead end}c). We let $S_{12} = 1000 \mu\textrm{m}^{2}$, $l_{12}  = 1 \textrm{mm}$,  $u_{12} = 1 \textrm{mm} \textrm{s}^{-1}$ and $u_{23} = 0$, while $D_{ij}$ is determined by Equation (\ref{definition_dispersion}). In this case we vary the length of the dead-end to see how it effects $C_{\textrm{tot}}$. As the presence of dead-ends vessels can only increase the mean transit time for resource crossing the network, we find that increasing the length of the dead-end regions increases the total rate of resource delivery (see Fig. \ref{delivery_dead_end}).
 
\subsection{Biomedical implications of altering vascular geometry}

Despite their simple nature we now suggest that the results of Section £.2 could have biomedical implications. Tumours require access to blood vessels for growth and metastasis. Consequently anti-angiogenic drugs, which disrupt and inhibit the formation of new blood vessels, are a promising avenue for the treatment of cancer. As single agents, anti-angiogenic drugs have only produced modest clinical improvements, but in combination with chemotherapy, the drug bevacizuab (a monoclonal antibody against vascular endothelial growth factor) has produced an unprecedented increase in survival (5 months) in colorectal cancer patients \cite{Hurwitz}. This is somewhat paradoxical, as previous studies have indicated that destroying the vasculature severely compromises the delivery of oxygen and therapeutics, producing hypoxia that renders chemotherapy and radiotherapy less effective \cite{Jain}. 

Tumours are subject to an unusually high interstitial fluid pressure, which may collapse blood and lymph vessels, and  inhibits the interstitial transport of drugs \cite{Truskey}. Furthermore, the blood vessels within tumours are relatively leaky, tortuous, and arranged in a haphazard, irregular pattern of interconnection, which results in velocities of fluid flow that vary spatially and temporally in a random manner \cite{Baish, Bullit, Jain, Pries, Truskey}. In healthy tissue the endothelial cells of the vasculature are supported by cells known as pericytes, but in tumours the pericytes are loosely attached or absent \cite{Carmeliet1, Jain, Winkler}. Anti-angiogenic drugs may impact on drug delivery in several ways:  they can induce the regression of the particularly leaky vessels that lack pericytes \cite{Jain}, they can encourage the maturation of the remaining vessels into less leaky, less dilated, less tortuous vessels with a greater coverage of pericytes \cite{Carmeliet1, Jain, Winkler}, they can alter the pattern of vascular adaptation \cite{Alarcon, Jain2, Owen} and they can reduce the interstitial fluid pressure \cite{Jain2, Shipley, Truskey}. Finally, by reducing the number of vessels, anti-angiogenic drugs alter the topology of the vasculature. These effects have been described as `vascular normalization' \cite{Baish, Jain}, and they help to explain why the use of anti-angiogenic drugs can actually increase the delivery capacity of the vasculature in tumours, increasing the chemo-sensitivity of the tumour itself.  
 
Our approach helps to illuminate the impact of changes in vascular geometry, as we can use our algorithm to compare the delivery rates of various substances for a pair of networks (before and after vascular pruning, say). If the delivery rate per unit of resource $R$ is small enough, almost every particle that enters the network will exit the network over a time-scale smaller than $1/R$. In that case the concentration of resource will be approximately constant throughout both networks. This implies that any reduction in the total volume of blood vessels will reduce the delivery capacity of the network for the substance in question, as the total rate of resource delivery is equal to the total volume of blood times the mean concentration of resource times $R$. On the other hand, if $R$ is sufficiently large, almost all the resource that enters the network will be consumed. Again we find that any reduction in the volume of blood vessels will reduce the total rate of resource delivery $C_{\textrm{tot}}$, but in this case it is because $C_{\textrm{tot}}$ is approximately equal to the current of resource entering the network, and reducing the number of blood vessels will increase the hydraulic resistance of the network, thereby reducing both the medium-current and the current of resource flowing into the network. 

The interesting case is also the most biologically relevant one: where $R$ is such that a significant amount of resource is present in the blood that is leaving the network, but the concentration of resource entering the network is significantly greater than the concentration of resource leaving the network. In this intermediate case, reducing the total volume of blood vessels may increase or decrease the delivery capacity of the network (that is, the total rate of resource delivery). If we ignore the impact of vascular pruning on interstitial pressure, Fig. \ref{delivery_dead_end} indicates that removing dead-ends can only reduce the delivery capacity of the vascular network. Essentially, removing such dead-end regions does not affect the amount of resource entering the network, but it does decrease the mean transit time. It follows that the resource flowing through the network is more likely to exit the network before it is consumed, which is to say that removing dead-end regions will decrease the delivery capacity of any given network. 

The effect of removing vessels that are an integral part of the network is more complex. In general, removing the shorter routes between the arteries and veins will increase the delivery capacity of the network, as, in the absence of short cuts, any resource that enters the network will be forced to spend longer within it, increasing the probability that any given particle will be consumed. As an extreme example, when an arteriovenous malformation is formed (that is, an abnormal connection between arteries and veins) the total volume of blood vessels increases, but such a malformation will effectively short-circuit the capillaries in the region, so the current in the capillaries and the rate of glucose and oxygen delivery drops dramatically \cite{Fleetwood}. As Fig. \ref{delivery_alternate_route} indicates, delivery is optimal when the various routes through the vasculature are of similar length, which indicates the importance of mechanisms that regulate the demarcation of artery-vein boundaries. This helps to explain the importance of Eph/ephrin signals, and other molecular cues that effectively identify endothelial cells as arterial or venous even before they are fused into a functioning circuit \cite{Augustin, Carmeliet2}. 

In conclusion, the effect of vascular pruning on glucose delivery will depend on the network structure, and the topological location of the vessels that are pruned. If anti-angiogenic drugs eliminate dead-end vessels, the treatment will decrease the mean transit time of blood flowing through the tumour. This will tend to reduce the total rate of glucose delivery and the chemo-sensitivity of the tumour (though this effect may be swamped by other effects of anti-angiogenic drugs, such as a reduction in interstitial pressure). On the other hand, if anti-angiogenic treatment eliminates the shorter routes by which blood transits through the tumour, our model suggests that the effect will be an increase in the total delivery rate of glucose, and an increase in the chemo-sensitivity of the tumour.

\section{Concentration in a growing fungal network} \label{fungal networks}

\subsection{Modelling the currents in fungal networks} \label{current in fungal network}

Multi-cellular organisms need to supply individual cells with the resources necessary for survival, but while transport in animals and plants is relatively well studied, surprisingly little is known about transport in the third major kingdom of multicellular life. The fungal body or mycelium can be understood as a network of fluid filled tubes or hyphae, which grow by osmotically drawing water from their surroundings while  adding material to the cell wall specifically at the tips of the growing hyphae \cite{Bartnicki, Money}. Diffusion may be sufficient to sustain short-range local growth when resources are abundant, but foraging fungi such as  \emph{Phanerochaete velutina} can grow hundreds of millimeters away from a food source over metabolically inert surfaces \cite{Bebber, Cairney, Olsson}. Together with various forms of experimental evidence, this observation strongly suggests that long-distance transport mechanisms are required to deliver nutrients to the growing tips at a sufficient rate, though there are many open questions concerning the mechanism(s) of transport \cite{Boswell, Cairney, Heaton, Jennings, Olsson}. Vesicles moved by motor proteins, contractile elements and carefully regulated osmotic gradients have all been proposed as mechanisms for driving long range transport in fungi \cite{Cairney, Jennings, Olsson}. Though a fundamental physiological question, which (if any) of these mechanisms is important remains debated.

We note that the fluid within fungal networks  is incompressible, and as the network grows, there is water uptake in and near the inoculum. It follows that there is a mass flow from the sites of water uptake to the sites of growth  \cite{Heaton}, and as the tips of the hyphae expand, the cytosol within the organism moves forward along with the growing tips \cite{Lew}. In this section we investigate the argument that this form of growth induced mass flow is sufficient to supply the growing tips with the resources they require. We do this by modelling advection, diffusion and delivery over empirically determined fungal networks.

To obtain a sequence of digitized fungal networks, we placed a woodblock inoculated with \emph{P. velutina} in a microcosm of compacted sand. The growing mycelium was photographed every three days, and the sequence of images was manually marked to record the location of nodes or junctions, as well as the presence or absence of edges in the fungal network. These edges were not sufficiently well resolved to make direct measurements of their diameter from the digitized images. However, the reflected intensity, averaged over a small user-defined kernel at either end of the edge, correlated well with microscope-based measurements of edge thickness. The observed relationship between image intensity and thickness was therefore used to estimate edge thickness across the whole network \cite{Bebber}.

The edges in our fungal networks are composed of bundles of hyphae and transport vessels bounded by an outer rind \cite{Eamus}. Unlike individual hyphae, the edges (or cords) in a fungal network have tough hydrophobic coatings which insulate them from the environment \cite{Cairney, Jennings}. We make two simplifying assumptions: we suppose that all the water and other materials which form the mycelium ultimately originate from the inoculum, and we suppose that each edge is composed of transport vessels, each of which has a typical radius of $6 \mu\textrm{m}$ \cite{Eamus}. Note that the latter of these assumptions implies that the hydraulic conductance of each edge is proportional to its cross-sectional area, as the number of transport vessels in each edge is proportional to its cross-sectional area. 

Since the mycelium is composed of incompressible material, the rate of increase in the volume of each edge must equal the volumetric rate of flow into that edge minus the volumetric rate of flow out of that edge. Together, these assumptions enable us to identify a unique medium-current for each edge, namely the set of medium-currents that are consistent with the observed changes in edge volume, and which also minimize the work required to overcome viscous drag \cite{Heaton}. In effect, we simply consider the mycelium as a network of resistors connecting the sources of material (the inoculum and shrinking edges) to sinks (the growing edges). This enables us to identify a minimal set of growth induced mass flows, and in this section we explore whether these currents are sufficient to deliver the resource that is required at the growing tips.

\subsection{Modelling resource uptake and delivery} \label{modelling uptake and delivery}

To find the distribution of resource that results from a given set of currents, we must make some assumptions about the rates of resource uptake and delivery. From the beginning of each experiment, the inoculum is filled with wood-degrading hyphae, so we assume that resource enters the network at the inoculum (node 1) at a constant rate $I_{1}(t) = K$. The rate of water uptake at the inoculum corresponds to the total rate of growth, so our assumptions imply that the amount of water entering the network per unit of resource is proportional to the rate of growth. We also suppose that throughout the network there is a constant rate of delivery per unit of resource $R$. In other words, where $Q(t)$ denotes the total amount of resource in the network, we suppose that
\begin{equation}
\frac{d}{dt}Q(t) = K - R Q(t).
\label{change in total resource}
\end{equation}
As $Q(0) = 0$, Equation (\ref{change in total resource}) implies that $Q(t) = \frac{K}{R} \big(1 -  e^{-Rt} \big)$.

The assumption that $K$ and $R$ are constants implies that the total quantity of resource in the network accumulates over a time-scale $\frac{1}{R}$, and approaches a steady state $\frac{K}{R}$. Furthermore, in our experimental set-up the fungal network attains a maximum volume as there is a finite quantity of resource for the fungi to consume. As a final assumption, we suppose that resource accumulates over a time-scale that is equal to the time-scale of growth, so that over the course of the experiment the mean concentration is approximately constant. More specifically, we let $V_{\textrm{F}}$ denote the maximum volume attained by the mycelium, and we measure the time $\tau$ that elapses before the mycelium attains a volume $\frac{1}{2} V_{\textrm{F}}$. We then assume that $Q(\tau)$ is half the maximum quantity of resource. The numerical value of $K$ reflects the units we use to measure the concentration, so without loss of generality we can assume that the mean concentration at time $\tau$ is 1. It follows that $Q(\tau) =  \frac{V_{\textrm{F}}}{2}$ and $\frac{K}{R} = V_{\textrm{F}}$. This implies that
\begin{equation}
R = \frac{\log (2)}{\tau} \quad \textrm{and} \quad K = \frac{V_{\textrm{F}} \log (2)}{\tau}, 
\label{k and R for growing fungi}
\end{equation}
so we have $Q(t) = V_{\textrm{F}} (1 -  2^{\frac{- t}{\tau}})$.

\subsection{Modelling the spatial distribution of resource in empirical networks} \label{empirical fungal networks}

To apply our minimal model for the distribution of resource in a growing fungal network, we require empirical values for $V_{\textrm{F}}$ (the maximum volume attained by the network) and $\tau$ (the time taken to grow to volume $\frac{1}{2} V_{\textrm{F}}$). We also require the adjacency matrix of the network, the lengths $l_{ij}$ and the cross-sectional areas $S_{ij}(t_{n})$ for each edge $ij$ and each time point $t_{1}, \ldots , t_{N}$.

For each time interval, the first step is to calculate the unique set of medium-currents which are consistent with the observed changes in volume, and which minimize the work required to overcome viscous drag \cite{Heaton}. We suppose that over the time interval $t_{n} < t \leq t_{n+1}$ the cross-sectional area $S_{ij}(t) = \frac{1}{2} \big(S_{ij}(t_{n}) + S_{ij}(t_{n+1}) \big)$. Furthermore, as the edges are composed of a bundle of transport vessels, we suppose that the conductance of each edge is proportional to its cross-sectional area. Finally, we calculate whether each of the nodes is a source or a sink. Where $F_{i}$ denotes the net medium-current flowing out of node $i$, we let
\begin{equation}
F_{i} = \Bigg\{
\begin{array}{ll}
-\sum_{j \neq i} F_{j} & \textrm{if node $i$ is the inoculum,} \\
& \\
\sum_{ij} \frac{S_{ij}(t_{n})-S_{ij}(t_{n+1})}{2(t_{n+1} - t_{n})}  & \textrm{otherwise.} \\
\end{array}
\label{q_def}
\end{equation}

If the edges around node $i$ are growing, then $F_{i}$ is negative and node $i$ is a sink, which is to say that more medium-current flows into the links of node $i$ than flow out. If the edges around node $i$ are shrinking, or if node $i$ is the inoculum, then $F_{i}$ is positive and node $i$ is a source. Circuit theory tells us that we can use the conductance of each edge and the net current flowing out of each node to determine the pressure difference between any pair of nodes \cite{Grimmett, Heaton}. Furthermore, given the conductance of edge $ij$ and the pressure drop between nodes $i$ and $j$, we can uniquely determine the medium-current $F_{ij}(t)$ for each edge in the network. This medium-current is constant over the time interval $t_{n} < t \leq t_{n+1}$, and it does not depend on the constant of proportionality between the cross-sectional area of the edges and the conductance of the edges. 


The edges or cords in a fungal network have a complex structure \cite{Eamus}, and mass flows occur in transport vessels that occupy some fraction $\lambda$ of the cross-sectional area of each edge. The medium-current in an edge is equal to the mean velocity of flow times the total cross-sectional area of the transport vessels, so for each edge and each time interval we have
\begin{equation}
u_{ij}(t) = \frac{2 F_{ij}(t)}{\lambda \big(S_{ij}(t_{n}) +  S_{ij}(t_{n+1}) \big)}.
\label{speed_in_cords}
\end{equation}

As we wish to investigate whether growth-induced mass flows are sufficient to carry resource from the inoculum to the tips over the time-scale of growth, we set $\lambda = 1$. In other words, given values for the medium-currents, we let the velocities of mass flow be as small as possible by maximizing $\lambda$. Also note that, given the observed changes in volume, and given the assumption that resource and water only enters the network at the inoculum, the medium-currents that we identify are as small as possible (in the sense that any other set of medium-currents consistent with the observed growth would require more work to overcome viscous drag). We are interested in finding the distribution of a generic source of energy and carbon, so we let $D_{m} = 6.7 \times 10^{-4} \textrm{mm}^{2}\textrm{s}^{-1}$ (the molecular diffusion coefficient of glucose), as this is representative of the diffusion coefficient of a small molecule. We also assume that in each edge the advection and diffusion of resource occurs within some number of transport vessels of radius $6 \mu\textrm{m}$ \cite{Eamus}, so once we have found the mean velocity of flow $u_{ij}(t)$, we can use Equation (\ref{definition_dispersion}) to find the piece-wise constant dispersion coefficient $D_{ij}(t)$. 

The delivery rate per unit of resource is assumed to be the same in every edge, and the value of $R_{ij} = R$ is given by Equation (\ref{k and R for growing fungi}). In each experiment the parameters $V_{\textrm{F}}$ and $\tau$ were chosen to ensure that, over each time step, the mean concentration is as close to 1 as possible. The nutrient and water content of woodblocks can vary, resulting in more or less vigorous growth. In the first replicate we found that  $V_{\textrm{F}} = 393 \textrm{mm}^{3}$ (20\% of the volume of the woodblock) and $\tau = 242 \textrm{ hours}$. In the second we found that $V_{\textrm{F}} = 372 \textrm{mm}^{3}$ (19\% of the volume of the woodblock) and $\tau = 468 \textrm{ hours}$. In the third we found that $V_{\textrm{F}} = 616 \textrm{mm}^{3}$ (31\% of the volume of the woodblock) and $\tau = 367 \textrm{ hours}$. Finally, the Laplace transform of the net quantity of resource leaving the inoculum is assumed to be $\Upsilon_{1}(s) = \frac{V_{\textrm{F}} \log (2)}{\tau s}$, and for every other node $\Upsilon_{i}(s) = 0$. We now have all the parameters we need to implement either of the algorithms described in the Appendix. That is to say, we can calculate the spatial distribution of resource that would arise if the cross-sectional areas of the edges vary in either a step-wise or continuous manner, where the volumetric currents are determined by the measured changes in volume. Whichever algorithm we we employ, at each time $t_{n}$ we record the resulting spatial distributions of resource by dividing each edge into $N_{ij}$ line segments such that $\frac{l_{ij}}{N_{ij}} \leq 1 \textrm{mm}$. These quantities of resource per unit length are then treated as an initial condition over the following time step, and the concentrations at time $t_{n}$ are identified by dividing $q_{ij}(x,t_{n})$ by $S_{ij}(t_{n})$.

\subsection{Results of the simulation}

\begin{figure}
\begin{center}
\includegraphics[width=12.7cm]{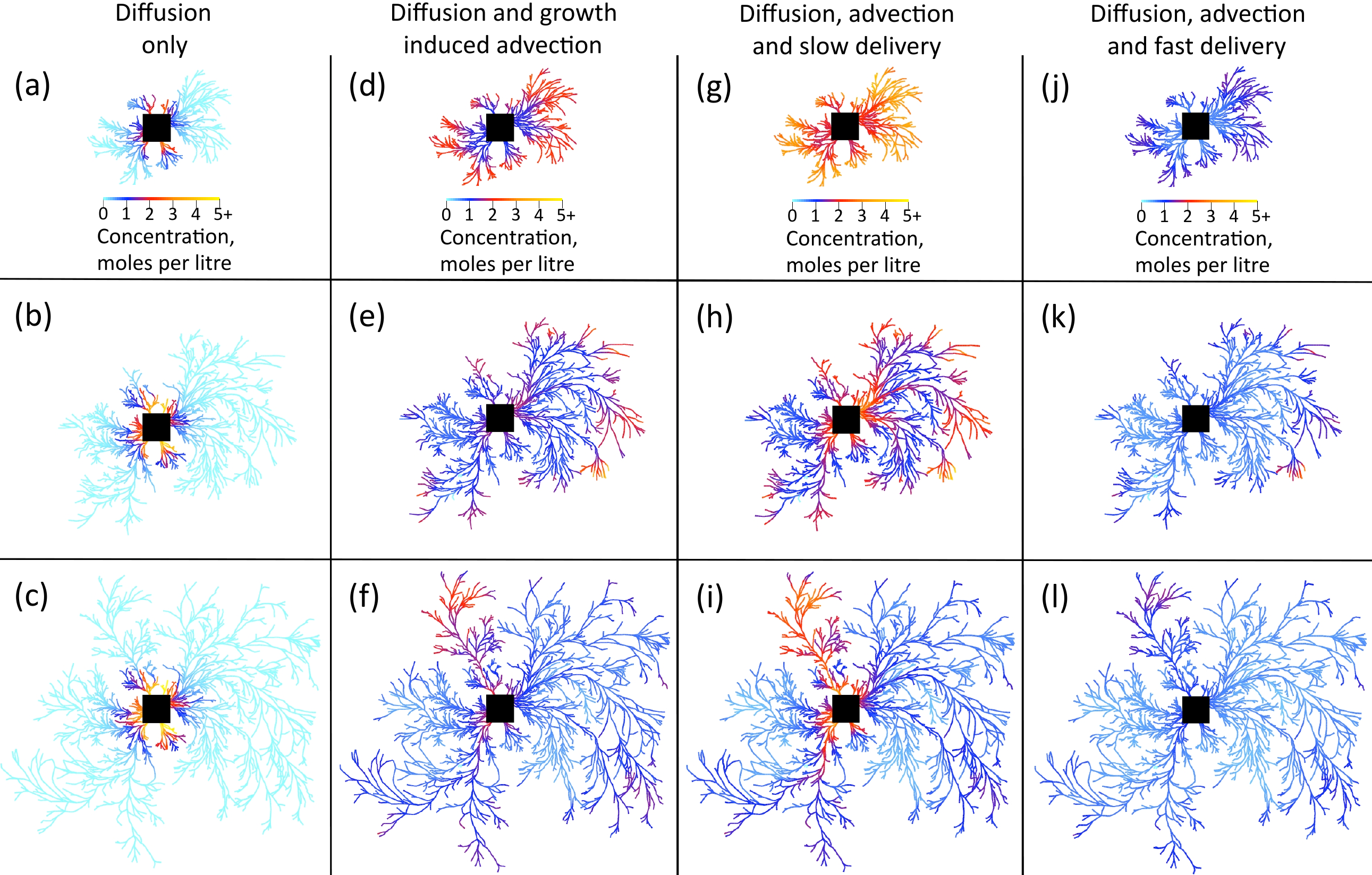}
\caption{\label{fungi_coloured_by_concentration} \textbf{Concentration of resource in Experiment 1.} Experiment 1 after 6 days (a, d, g and j), 12 days (b, e, h and k) and 18 days (c, f, i and l). Diagrams (a - c) illustrate the concentration of resource that would occur in the absence of advection, where resource enters at the inoculum at a rate of $K = 1.125 \mu\textrm{mole hour}^{-1}$, and $\tau = 242 \textrm{ hours}$. Diagrams (d - f) illustrate the concentration of resource where fluid and resource enter the network at the inoculum, and the medium-currents are consistent with the observed changes in volume, while minimising the work required to overcome viscous drag (see Section \ref{current in fungal network}). As before, resource enters at the inoculum at a rate of $K = 1.125 \mu\textrm{mole hour}^{-1}$, and $\tau = 242 \textrm{ hours}$. Note that at any point in time, the concentration near the tips can be greater than the concentration near the inoculum. This is possible because resource enters the network at a constant rate, but the rate of water influx at the inoculum corresponds to the total volumetric rate of growth. Consequently, as the total volumetric rate of growth increases, the concentration of resource in the fluid near the inoculum decreases. In (d), for example, the fluid in the tips contains more resource than the fluid near the inoculum, but when that fluid first entered the network (at the inoculum) it contained an even higher concentration of resource. We cannot directly measure the delivery rate of resource, so to assess the sensitivity of our model to the parameter $R$, we also consider the cases where we half and double the delivery rate $R$. Diagrams (g - i) illustrate the concentrations that occur when the medium-currents and rate of uptake are as before, but the local delivery rate has been halved. Diagrams (j - l) illustrate the concentrations that occur when the medium-currents and rate of uptake are as before, but where the local delivery rate has been doubled.}
\end{center}
\end{figure}

\begin{figure}
\begin{center}
\includegraphics[width=7.5cm]{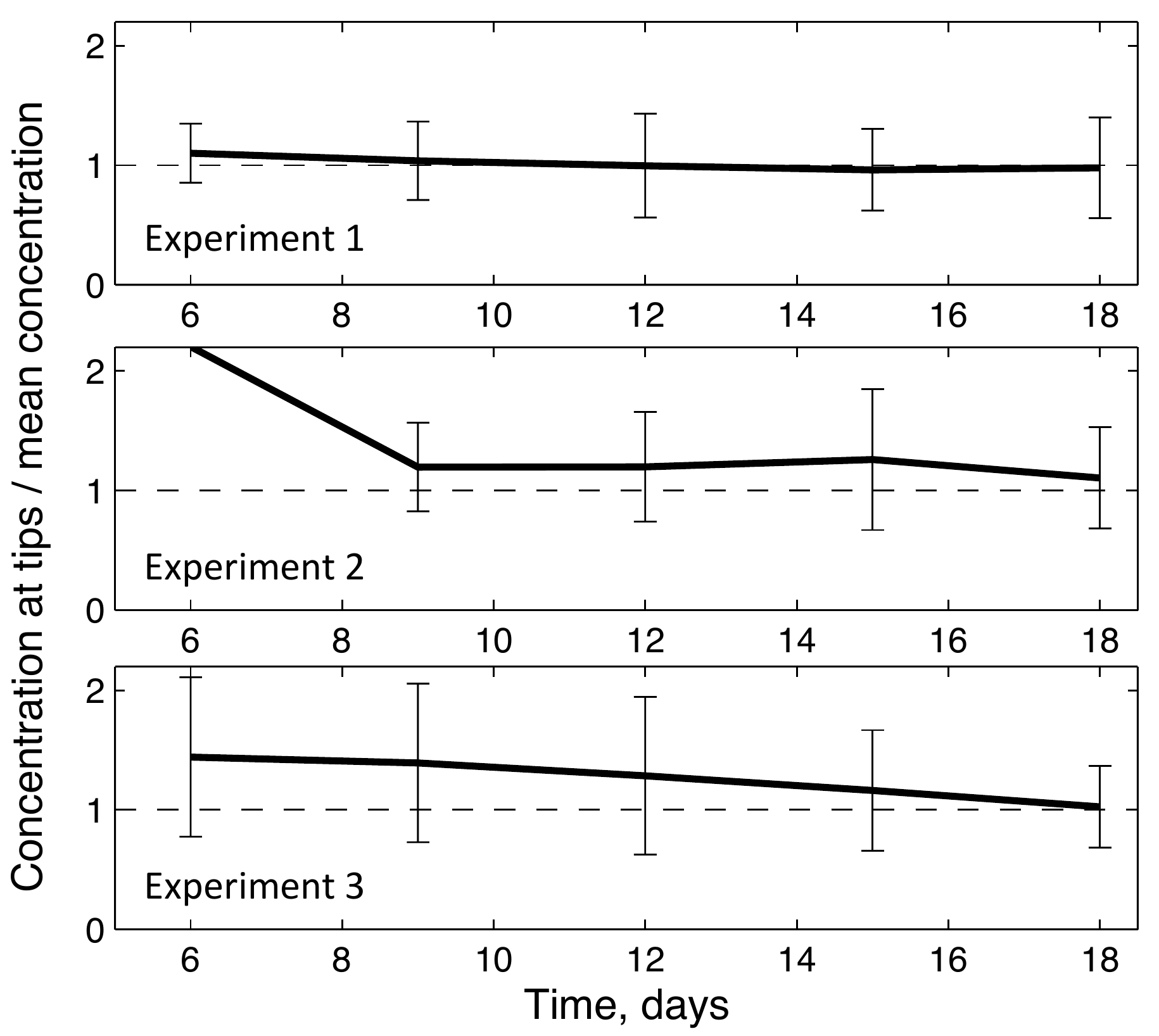}
\caption{\label{ratio_of_conc_at_tips_to_mean}  \textbf{Concentration in the tips relative to the concentration elsewhere.} In each of the three experiments, the concentration at the tips (nodes of order one) was larger than the mean concentration in the network as a whole.}
\end{center}
\end{figure}

Three fungal networks were grown and digitized, and the observed changes in fungal volume were used to determine the minimal currents consistent with the changing volume, as well as the uptake rate and decay rate of a generic form of resource (see Sections \ref{current in fungal network}, \ref{modelling uptake and delivery} and \ref{empirical fungal networks}). In each of three experiments our model suggests that the growth induced mass flows were sufficiently large to spread the resource from the inoculum out to the growing tips over the time-scale of growth (see Figs. \ref{fungi_coloured_by_concentration} and \ref{ratio_of_conc_at_tips_to_mean}).

This result is somewhat counter-intuitive, as in most of the edges the mean velocity of the growth induced mass flows is very low \cite{Heaton}. Indeed, if we pool the data from all three experiments and over all time steps,  70\%  of the edges have a mean velocity that is so small that over the course of one week, resource travelling at that velocity would move \emph{less} than the $20 \textrm{mm}$ that resource would typically travel by diffusing in one dimension (75\% of edges in Experiment 1,  59\% in Experiment 2 and  64\% in Experiment 3). Over the time-scale of two hours, only 4\% of edges have a velocity great enough to carry resource further than the $2.2 \textrm{mm}$ that is typically travelled by diffusion alone (2\% in Experiment 1,  6\% in Experiment 2 and  6\% in Experiment 3).

Despite the modest scale of the advection in most of the edges, the fraction of edges in which the mean velocity is significant suffices to spread the resource from the inoculum out to the growing tips (see Figs. \ref{fungi_coloured_by_concentration} and \ref{ratio_of_conc_at_tips_to_mean}). We also calculated the distribution of resource that results if the cross-sectional areas $S_{ij}(t)$ vary continuously over each time step (see AIV), but the results were almost identical to the simpler case where the cross-sections are varied in a stepwise manner.

\subsection{Discussion of growth induced mass flows} 

\begin{figure}[t!]
\begin{center}
\includegraphics[width=8.5cm]{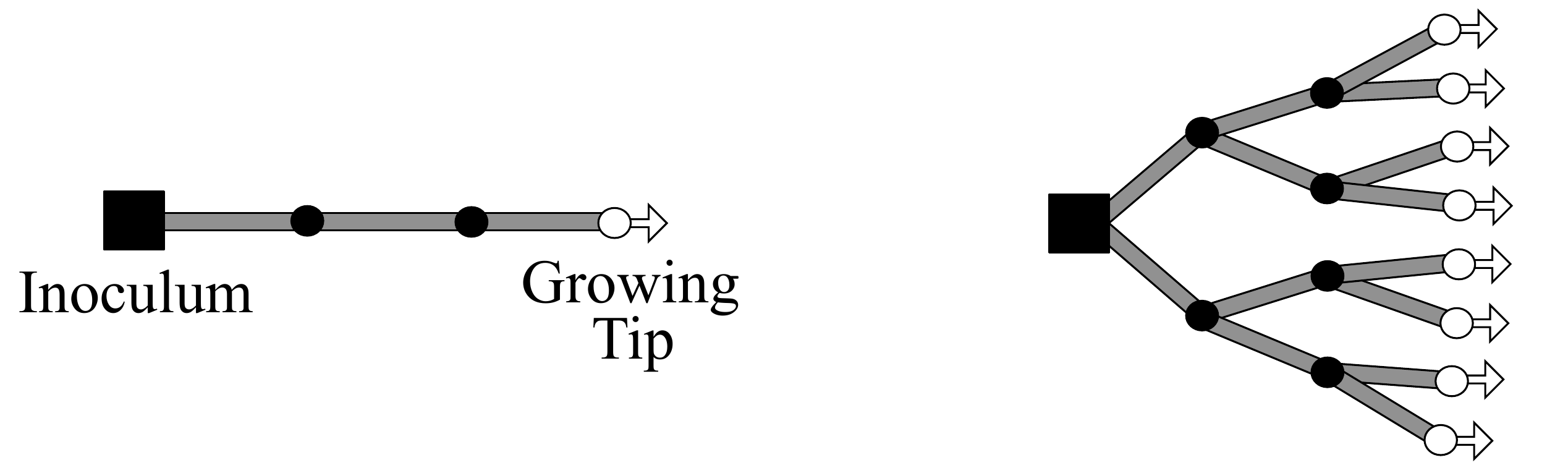}
\caption{\label{linear_network_and_branching_tree}  \textbf{Network structure and the efficacy of growth induced mass flows as a means of transport.} Two contrasting examples of networks with growth induced mass flows.}
\end{center}
\end{figure}

In a fungal network, the incompressibility of aqueous fluids ensures that growth in one part of the network requires the presence of fluid flows in the supporting mycelium. By controlling the spatial location of growth, maintaining the appropriate turgor pressure and by forming cords that are insulated from the environment, fungi can ensure that there is a long range flow of fluid from the sites of water uptake to the sites of growth \cite{Heaton}. Furthermore, the structure of the network is critical for ensuring that growth induced mass flows can carry resource from the inoculum to the growing tips over a reasonable period of time. In this regard, it is instructive to compare a growing linear network to a growing branching tree (see Fig. \ref{linear_network_and_branching_tree}). 

Suppose that the tips in both of the model networks illustrated in Fig. \ref{linear_network_and_branching_tree} grow a unit distance from the inoculum per unit time, and that each edge has unit length and volume. In the case of a linear network, and in the absence of diffusion, it will take $n$ units of time for the resource to travel from the inoculum to the $n$th edge. It follows that if the time-scale of delivery is $n$, then growth induced mass flows in a linear network cannot supply resource over length scales greater than $n$. In the case of a branching tree the volume of the $n$th generation is greater than the total volume of all the preceding generations (see Fig. \ref{linear_network_and_branching_tree}), so it will take less than a unit of time for resource to travel from the inoculum to the $n$th generation. Provided that the concentration of resource at the inoculum remains sufficiently high, there is no limit to the size of branching tree that can be filled with resource by growth induced mass flows, even if the local rate of resource delivery is high. Also note that in the absence of diffusion, the fluid exactly at the growing tips is never replaced by the fluid entering at the inoculum. This implies that growth induced mass flows alone cannot supply resource to the growing tips: diffusion and specific transport mechanisms are essential for transporting resource across the newest generation of edges.

Returning to our model of transport and resource delivery in an empirical fungal network, it could be argued that the total rate of resource delivery in an edge should be proportional to the volume of that edge, rather than being proportional to the quantity of resource contained within that edge. However, as our model results in a fairly constant concentration throughout the network, changing from first order to zero order delivery is unlikely to make a significance difference to the concentration at the tips (unless, of course, we change the mean amount of time that elapses between resource entering the network and the resource being consumed). It might also be argued that the growing hyphal tips are responsible for a significant fraction of the resource consumption \cite{Boswell, Olsson}. After all, as material is added to the growing cell wall, the concentration of that material in the cytoplasm must be depleted near the region of growth. Although our model does not include a term for consumption due to growth, it does indicate that growth induced mass flows are sufficient to carry resource across the network over the time-scale of growth. Furthermore, Equation (\ref{speed_in_cords}) indicates that if the growth induced mass flows are confined to transport vessels that only occupy a fraction $\lambda$ of the total cross section of each edge, then the mean advective velocities will be greater than our minimal estimates by a factor of $1/\lambda$.  

While advective mass flows carry resource over long distances from the inoculum out towards the growing tips \cite{Cairney, Jennings, Olsson}, diffusion and active transport mechanisms may be essential near the sites where the cell wall is  expanding. This follows because the cytosol within the hyphae moves forward at the same rate as the growing tips \cite{Lew}, but to transport resource from the base of the hyphae to the growing tips, the resource has to move faster than the rate of growth. Complex cellular machinery regulates the addition of material to the cell walls, ensuring that the growing hyphae exhibit polar growth, and only expand at the hyphal tips \cite{Bartnicki, Money}. We note, however, that our model suggests that vesicles carried by motor proteins or other active transport mechanisms may not be needed for longer range transport within fungal networks. 

\subsection*{Conclusion}
In the  Appendix we present two algorithms for calculating the concentration of resource that arises when a given material is subject to advection, diffusion and local delivery out of the transport network. The resulting spatial distribution will depend on the time-scale required to transit the network, and the time-scale of delivery (see Section \ref{time-scales and transport}). Nature is full of networks in which materials within a fluid are transported by advection and diffusion while being consumed or delivered, so these algorithms have many potential applications. In particular, our modelling framework can be applied to the case of glucose delivery through a vascular network. By analyzing simple, idealized vascular networks we found that the total rate of glucose delivery depends on the network structure (see Section \ref{idealized vascular networks}), and in some cases increasing the volume of blood and the number of glucose transporters can actually decrease the total rate of glucose delivery. This counter-intuitive result can occur because the additional vessels can decrease the time taken to transit the network, allowing a greater fraction of the glucose to pass through the network without encountering a transporter. Finally, we employed our algorithms to implement a model of transport in a growing fungal network (see Section \ref{fungal networks}). The expansion of fluid filled vessels requires the movement of fluid, and in three empirically determined fungal networks we found that the minimum currents consistent with the observed growth would effectively transport resource from the inoculum to the growing tips over the time-scale of growth. This suggests that the active transport mechanisms observed in the growing tips of fungal networks may not be required for long range transport.

\subsection*{Acknowledgements}
LLMH thanks the EPSRC for financial support, and Sid Redner for his helpful comments. EL thanks the grant EP/E056997/1. PKM was partially supported by a Royal Society Wolfson Research Merit Award. MDF thanks the BBSRC and NERC. NSJ thanks grant numbers EP/I005765/1 and EP/I005986/1.

\newpage
\begin{center}
\huge{Appendix: Mathematical methods.}
\end{center}
\section{Advection, diffusion and delivery in Laplace space} \label{advection etc in Laplace space}

As we explain in the Main Text, we are motivated to consider the case where resource is lost or delivered out of a given network at some local rate, while the resource that remains within the network moves by advection and diffusion. Such a process will result in a spatial distribution of resource that changes over time. We only consider longitudinal coordinates along each edge $ij$, using real numbers $x$ to denote distances from node $i$, where $0 \leq x \leq l_{ij}$. Each edge contains a quantity of resource, which must satisfy the one-dimensional advection-diffusion-delivery equation
\begin{equation}
\frac{\partial q_{ij}}{\partial t} + R_{ij} q_{ij} + u_{ij} \frac{\partial q_{ij}}{\partial x} - D_{ij} \frac{\partial^{2} q_{ij}}{\partial x^{2}} = 0,
\label{basic_CDE}
\end{equation}
where $q_{ij}$ is the quantity of resource per unit length, $u_{ij}$ is the mean velocity, $D_{ij}$ is the dispersion coefficient and $R_{ij}$ is the rate at which a unit of resource is lost, or delivered out of the network. As we explain in the Main Text Section 2.2, Fick's law implies that
\begin{equation}
I_{i}(t) = \sum_{j} \bigg[ u_{ij}(t) q_{ij}(x,t) - D_{ij} \frac{\partial q_{ij}(x,t)}{\partial x} \bigg]_{x=0},
\label{sums_of_J}
\end{equation}
where $I_{i}(t)$ denotes the net quantity of resource leaving node $i$ at time $t$. Note that by the conservation of mass any resource that enters node $i$ must leave node $i$ (as nodes have zero volume), so $I_{i}(t) = 0$ unless resource is entering the network at node $i$. Nodes where resource enters the network are referred to as inlet nodes.

Given such a system of fundamental equations, we want to find the quantity of resource throughout the network in an efficient manner. It is convenient to follow the approach of [39], which exploits the Laplace transform. This operation effectively weights the different time-scales over which resource may move from one node to another, and it is an efficient way of coping with the wide range of velocities that our network may contain. In particular, we take advantage of the following properties of the Laplace transform $\mathcal{L}\big( q_{ij}(x,t) \big) = \int_{0}^{\infty} q_{ij}(x,t)e^{-st}dt = Q_{ij}(x,s)$:
\begin{equation}
\mathcal{L}(\frac{\partial q_{ij}(x,t)}{\partial t}) = s Q_{ij}(x,s) - q_{ij}(x, 0), \qquad \textrm{and}
\label{laplace_of_time_derivative}
\end{equation}
\begin{equation}
\mathcal{L}(\frac{\partial q_{ij}(x,t)}{\partial x}) = \frac{\partial}{\partial x} Q_{ij}(x,s).
\label{laplace_of_space_derivative}
\end{equation}

If $u_{ij}$ and $D_{ij}$ are constant over time, it follows from Equation (\ref{basic_CDE}) that
\begin{equation} 
(s + R_{ij})Q_{ij} + u_{ij} \frac{\partial Q_{ij}}{\partial x} - D_{ij} \frac{\partial^{2} Q_{ij}}{\partial x^{2}} = q_{ij}(x,0).
\label{laplace_of_CDC_equation}
\end{equation}
Furthermore, Equations (\ref{sums_of_J}), (\ref{laplace_of_time_derivative}) and  (\ref{laplace_of_space_derivative}) imply that
\begin{equation}
\Upsilon_{i}(s) = \sum_{j} \bigg[ u_{ij} Q_{ij}(x, s) - D_{ij} \frac{\partial}{\partial x}Q_{ij}(x, s) \bigg]_{x=0},
\label{laplace_of_J}
\end{equation}
where $\Upsilon_{i}(s)$ denotes the Laplace transform of $I_{i}(t)$. 

\subsection{Zero initial conditions in an edge} \label{zero initial conditions}
We begin by considering an initially empty edge, before extending our results to the more complex case of non-zero initial conditions. We let  $q_{ij}(x, 0)=0$, and consider the homogenous case of Equation (\ref{laplace_of_CDC_equation}):
\begin{equation}
(s + R_{ij}) Q_{ij} + u_{ij}\frac{\partial Q_{ij}}{\partial x} - D_{ij} \frac{\partial^{2} Q_{ij}}{\partial x^{2}} = 0.
\label{homogenous_laplace_of_CDC}
\end{equation}
By solving this ODE in the usual manner, we find that for some pair of constants $A$ and $B$,
\begin{equation}
Q_{ij}(x,s)= Ae^{\frac{u_{ij} + \alpha_{ij}(s)}{2D_{ij}} x} + Be^{\frac{u_{ij} - \alpha_{ij}(s)}{2D_{ij}} x}, 
\label{Q_in_terms_of_A_and_B}
\end{equation}
\begin{equation}
\textrm{where} \quad \alpha_{ij}(s) =  \sqrt{u_{ij}^{2} + 4D_{ij}(s + R_{ij})}.
\label{definition_alpha}
\end{equation}
Note that the Laplace variable $s$ represents a rate, and that $\alpha_{ij}(s) = \alpha_{ji}(s)$ is positive, and dimensionally equivalent to speed. Roughly speaking, $\alpha_{ij}(s)$ represents the speed at which resource travels over the time scale $1/s$, with a correction term to account for delivery. Since $s$ and $D_{ij}$ are positive and $R_{ij}$ is non-negative, we always find that $\alpha_{ij}(s) > \big| u_{ij} \big|$. When $s = \frac{u_{ij}^{2}}{4 D_{ij}} - R_{ij}$, Equation (\ref{definition_alpha}) implies that $\alpha_{ij}(s) = \sqrt{2 u_{ij}^{2}}$. The value of $\alpha_{ij}(s)$ depends on $u_{ij}$, $D_{ij}$ and $R_{ij}$ over most time scales, but for very short time scales ($s \gg \frac{u_{ij}^{2}}{4 D_{ij}} - R_{ij}$) almost all the movement is due to diffusion,  $\alpha_{ij} \gg u_{ij}$ and $\alpha_{ij}(s) \approx \sqrt{4 D_{ij} (s + R_{ij})}$.

Equation (\ref{Q_in_terms_of_A_and_B}) tells us that for any positive number $s$, we can find $A$ and $B$ and express $Q_{ij}(x,s)$ in terms of the quantity of resource at either end of the edge. For any given $s$, we denote the quantity of resource at the ends of each edge by 
\begin{equation}
X_{ij}(s) \equiv Q_{ij}(0,s) \qquad \textrm{and} \qquad X_{ji}(s) \equiv  Q_{ji}(0,s) = Q_{ij}(l_{ij},s).
\label{defn_Xij_and_Xji}
\end{equation}  
For each edge $ij$, it is convenient to define two, dimensionless ratios between time scales:
\begin{equation}
g_{ij} = \frac{u_{ij} l_{ij}}{2 D_{ij}} \quad \textrm{ and } \quad h_{ij}(s) = \frac{\alpha_{ij}(s) l_{ij}}{2 D_{ij}}.
\label{definition_g_and_h}
\end{equation}
Setting $x = 0$ and $x = l_{ij}$ tells us that
\begin{equation}
X_{ij} = A + B \qquad \textrm{and} \qquad X_{ji} = Ae^{(g_{ij} + h_{ij})} + B e^{(g_{ij} - h_{ij})}.
\label{Qij_and_Qji}
\end{equation}

$A$, $B$, $X_{ij}$, $X_{ji}$, $\alpha_{ij}$ and $h_{ij}$ are all functions of the Laplace variable $s$, but this dependence is omitted for the sake of clarity. Equation (\ref{Qij_and_Qji}) tells us that
\begin{equation*}
Ae^{(g_{ij} + h_{ij})} + (X_{ij} - A)e^{(g_{ij} - h_{ij})} = X_{ji}, \quad \textrm{so} 
\end{equation*}
\begin{equation}
A = \frac{X_{ji}e^{-g_{ij}} - X_{ij}e^{-h_{ij}}}{2  \textrm{ sinh}(h_{ij})} \quad \textrm{and likewise}
\label{definition_A_zero_initial_conditions}
\end{equation}
\begin{equation}
B = \frac{X_{ij}e^{h_{ij}} - X_{ji}e^{-g_{ij} }}{2  \textrm{ sinh}(h_{ij})} .
\label{definition_B_zero_initial_conditions}
\end{equation}

Note that if $u_{ij}$ is negative, the medium-current flows towards node $i$ and the macroscopic P\'{e}clet number for the edge $ij$ is $\frac{ u_{ji} l_{ij}}{D_{ij}} = -2g_{ij}$ [39, 60]. Assuming that edge $ij$ is initially empty, we can find $Q_{ij}(x,s)$ by substituting Equations (\ref{definition_g_and_h}), (\ref{definition_A_zero_initial_conditions}) and  (\ref{definition_B_zero_initial_conditions}) into Equation (\ref{Q_in_terms_of_A_and_B}), giving us
\begin{eqnarray}
Q_{ij}(x,s) 
& = & X_{ij} \frac{  \textrm{ sinh}(\frac{l_{ij} - x}{l_{ij}}h_{ij}) }{  \textrm{ sinh}(h_{ij}) } e^{\frac{x}{l_{ij}}g_{ij}} +  X_{ji}   \frac{  \textrm{ sinh}(\frac{x}{l_{ij}}h_{ij}) }{  \textrm{ sinh}(h_{ij}) } e^{\frac{x - l_{ij} }{l_{ij}} g_{ij}}. 
\label{Cij_as_a_function_of_x}
\end{eqnarray}

\subsection{Advection, diffusion and delivery in an initially empty, static network} \label{system of equations}
Having examined the case of a single edge, we now turn to the problem of coupling the edges of a network such that the concentrations vary continuously as we move from one edge to another. For each node $i$ we have $C_{i}(s) = \int_{0}^{\infty} c_{i}(t)e^{-st}dt$. Assuming that the cross-sectional areas $S_{ij}$ are constant, Equations (Main Text 3),  and (\ref{defn_Xij_and_Xji}) imply that for all edges $ij$ we have
\begin{equation}
C_{i}(s) = \frac{X_{ij}(s)}{S_{ij}}  \quad \textrm{and} \quad C_{j}(s) = \frac{X_{ji}(s)}{S_{ij}}.
\label{definition_Ci}
\end{equation}
Enforcing this equation ensures that the Laplace transform of the concentration at node $i$ is consistent for all edges $ij$, $ik$, and so on. In general, we may not know the Laplace transform of the node concentrations $\bar{C}(s) = \{C_{1}(s), \ldots , C_{m}(s) \}$, where $m$ is the number of nodes. However, given $\bar{I}(t) = \{ I_{1}(t), \ldots , I_{m}(t) \}$ (the net current of resource leaving each node), we can calculate $\bar{\Upsilon}(s) = \{ \Upsilon_{1}(s), \ldots , \Upsilon_{m}(s) \} $ (the Laplace transform of $\bar{I}$), and, in the following manner, calculate $\bar{C}(s)$. If we substitute Equation (\ref{Q_in_terms_of_A_and_B})  into Equation (\ref{laplace_of_J}), noting that $x = 0$ tells us that
\begin{eqnarray*}
\Upsilon_{i}(s) 
  & = & \sum_{j}  \frac{\alpha_{ij}}{2} \big(B - A \big) + \frac{u_{ij}}{2} \big(A + B\big).  \nonumber \\
  \end{eqnarray*}
Equations (\ref{definition_g_and_h}) and (\ref{Qij_and_Qji}) imply that $A + B = X_{ij}$, and
\begin{eqnarray*}
B - A & = & \frac{1}{2  \textrm{ sinh}(h_{ij})} \bigg[ X_{ij}e^{h_{ij}} - X_{ji}e^{-g_{ij}} - X_{ji}e^{-g_{ij}} + X_{ij}e^{-h_{ij}}  \bigg], 
\end{eqnarray*}
so we have
\begin{eqnarray}
\Upsilon_{i}(s) & = & \sum_{j}  \bigg[ \frac{\alpha_{ij}}{2  \textrm{ sinh}(h_{ij})} \bigg( X_{ij}  \textrm{ cosh}(h_{ij})  -  X_{ji}e^{-g_{ij}} \bigg) + \frac{u_{ij}}{2} X_{ij}  \bigg].
\label{upsilon_in_terms_of_Xij_Xji}
\end{eqnarray}

Equations (\ref{definition_Ci}) and (\ref{upsilon_in_terms_of_Xij_Xji}) imply that
\begin{eqnarray}
\Upsilon_{i}(s) & = & \sum_{j} \bigg[ C_{i}(s)  S_{ij} \bigg( \frac{u_{ij}}{2} + \frac{\alpha_{ij}}{2  \textrm{ tanh}(h_{ij})} \bigg)  -  C_{j}(s)  S_{ij} \bigg( \frac{\alpha_{ij} e^{-g_{ij}}}{2  \textrm{ sinh}(h_{ij})} \bigg) \bigg]. \qquad
\end{eqnarray}
\\
In other words, for each node $i$ we have a linear equation in $C_{1}(s), C_{2}(s), \ldots, C_{m}(s)$. Hence where $\bar{C}(s)$ and $\bar{\Upsilon}(s)$ are column vectors, we thus have
\begin{equation}
\textbf{M}(s)\bar{C}(s) = \bar{\Upsilon}(s),
\label{matrix_M_for_CDE}
\end{equation}
where 
\begin{equation}
\textbf{M}_{ij}(s) =  \Bigg\{	
\begin{array}{cl}
\sum_{k} S_{ik} \bigg[ \frac{u_{ik}}{2} + \frac{\alpha_{ik}}{2  \textrm{ tanh} ( h_{ik})} \bigg] & \textrm{ if $i = j$,} \\
& \\
\frac{- S_{ij} \alpha_{ij} e^{-g_{ij}}}{2  \textrm{ sinh}\big( h_{ij} \big)} & \textrm{ otherwise.}  
\end{array}
\label{defn_M}
\end{equation}

We refer to the matrix $\textbf{M}(s)$ as the propagation matrix, and it contains a row and column for each node in the given network. Given  $\textbf{M}(s)$ and $\bar{\Upsilon}(s)$ we can calculate $\bar{C}(s)$ using various efficient algorithms, including the stabilized biconjugate gradient method (BiCGStab). In most cases this is the most efficient algorithm to use, as our matrix $\textbf{M}(s)$ is non-symmetric and sparse [24]. 

Equation (\ref{definition_alpha}) implies that the diagonal elements $\mathbf{M}(s)$ are all positive. Furthermore, $\mathbf{M}_{ij}(s) = 0$ if and only if there is no edge between $i$ and $j$, and the other off-diagonal elements are negative. We note that if there is resource at node $j$, it may be transported along $ij$, bringing resource to $i$ and reducing $\Upsilon_{i}(s)$ (the Laplace transform of the net current flowing out of node $i$). Resource can only reach node $i$ along the edges $ij$, $ik$, etc, so $\Upsilon_{i}(s)$ is completely determined by the concentration at $i$ and the concentrations that flow through the nodes adjacent to $i$. As $\Upsilon_{i}(s)$ is the Laplace transform of the net current flowing out of node $i$, and resource at nodes $j \neq i$ can flow into node $i$, the off-diagonal elements of $\mathbf{M}(s)$ are negative, and zero if $i$ and $j$ are not directly connected.

Multiplying $\big| \mathbf{M}_{ij}(s) \big|$ by $C_{j}(s)$ gives us the Laplace transform of the current of resource flowing from $j$ to $i$, so roughly speaking, $\big| \mathbf{M}_{ij}(s) \big|$ represents the size of the volumetric current from $j$ to $i$, over the time scale $1/s$. Note that if $u_{ij}$ is positive, then the medium-current flows from $i$ to $j$, $\big| \mathbf{M}_{ij}(s) \big| < \big| \mathbf{M}_{ji}(s) \big|$, and there is a greater flow from $i$ to $j$ than the other way around. That is to say, when the medium-current is from $i$ to $j$, the value of $C_{i}(s)$ has a greater influence on the value of $\Upsilon_{j}(s)$ than the influence of  $C_{j}(s)$ on the  value of $\Upsilon_{i}(s)$. Also note that the ratio of $\mathbf{M}_{ij}(s)$ to $\mathbf{M}_{ji}(s)$ depends on the P\'{e}clet number $\frac{u_{ij} l_{ij} }{D_{ij} } = 2 g_{ij}$, as $\mathbf{M}_{ij}(s):\mathbf{M}_{ji}(s)$ is equal to $1:e^{2 g_{ij}}$.

For very short time scales we have a very large $s$, and by Equation (\ref{definition_alpha}), $\alpha_{ij} \gg u_{ij}$ and $\alpha_{ij} \approx \sqrt{4 D_{ij} s}$. In this case the off-diagonal elements of $\mathbf{M}$ are very small, and $\mathbf{M}_{ii} \approx \sum_{k} S_{ik} \frac{\alpha_{ik}}{2} \approx \sum_{k} S_{ik} \sqrt{D_{ik} s}$. In other words, over very short time scales resource is lost from the nodes by a process of diffusion, but it does not have time to reach the other nodes. Over longer time scales the difference between $u_{ij}$ and $\alpha_{ij}$ is smaller, the off-diagonal elements of $\mathbf{M}$ are larger, and effect of advection is greater.



\subsection{Inverting from Laplace space} \label{inverting from Laplace space}
We now have a method for finding the Laplace transform of various quantities, and in this section we consider how to transform these quantities into the time domain. More specifically, we have seen that for a given Laplace value $s$, we can find $\textbf{M}(s)$ and $\bar{\Upsilon}(s)$. We can therefore use Equation (\ref{matrix_M_for_CDE}) to find $\bar{C}(s) = \{C_{1}(s), \ldots , C_{m}(s) \}$, the Laplace transform of the concentrations at each node. Furthermore, we can use Equation (\ref{Cij_as_a_function_of_x}) to calculate $Q_{ij}(x,s)$ in terms of the boundary conditions $X_{ij}(s) = \mathcal{L}\big( q_{ij}(0, t) \big)$ and  $X_{ji}(s) = \mathcal{L}\big( q_{ij}(l_{ij}, t) \big)$. In other words, for each edge and each Laplace variable $s$, we can find an algebraic expression for $Q_{ij}(x,s)$ in terms of the boundary conditions $X_{ij}(s)$ and $X_{ji}(s)$, but we have yet to show how we can numerically invert such quantities into the time domain.  

As we can calculate any sequence of real valued sample points in Laplace space and we wish to calculate the corresponding value at a given point in time, it is appropriate and efficient to apply the Gaver-Stehfest algorithm [2, 22]. The key idea behind this algorithm (and other, related algorithms) is the notion of constructing a sequence of linear combinations of exponential functions, in order to form a weighted delta convergent sequence [2, 22, 65, 66]. That is to say, we consider a sequence of functions $\delta_{n}(x, t)$ such that for any function $q$ that is continuous at $t$, we have
\begin{equation}
\int_{0}^{\infty} \delta_{n}(v, t) q(v) dv = t \tilde{q}_{n}(t),
\label{defn_delta}
\end{equation}
where $\tilde{q}_{n}(t) \rightarrow q(t)$ as $n \rightarrow \infty$. As we shall see, there are weighted delta convergent sequences of functions such that $\delta_{n}(v, t)$ is of the form 
\begin{equation}
\delta_{n}(v, t) = \sum_{i=1}^{n} \omega_{i} e^{\frac{- \theta_{i} v}{t} },
\label{form_delta}
\end{equation}
where $\theta_{i} > 0$ for all $i$, and the terms $\theta_{i}$ and $\omega_{i}$ do not depend on $t$. Now, if we suppose that our function $q$ does not increase exponentially, then the Laplace transform $Q(s) = \int_{0}^{\infty} e^{-s v} q(v) dv$ is well defined for all positive numbers $s$. Hence the existence of $Q(s)$ for all positive $s$ is a reasonable assumption, given the context in which our functions $q$ arise. Assuming that $Q(s)$ is well defined for all positive numbers $s$, Equations (\ref{defn_delta}) and (\ref{form_delta}) imply that
\begin{eqnarray*}
\tilde{q}_{n}(t) & = & \frac{1}{t} \int_{0}^{\infty} \sum_{i=1}^{n} \omega_{i} e^{\frac{- \theta_{i} v}{t} } q(v) dv  = \frac{1}{t} \sum_{i=1}^{n} \omega_{i} Q \big(\frac{\theta_{i}}{t} \big).
\end{eqnarray*}

Gaver [28] employed the sequence of functions
\begin{equation*}
\delta_{n}(v, t) = \ln 2 \frac{(2n)! }{n!(n-1)!} (1 - e^{-\frac{v \ln 2}{t} } )^{n} (e^{-\frac{v \ln 2}{t} } )^{n},
\end{equation*}
but the resulting terms $\tilde{q}^{n}(t)$ converge to $q(t)$ logarithmically slowly. Gaver also showed that the quantity $\tilde{q}^{n}(t) - q(t)$ can be expanded in terms of inverse powers of $n$, which enabled him to accelerate the convergence of his original sequence of approximations [28]. The most useful formula for finding an accurate estimate of $q(t)$ based on a linear combination of the Gaver estimates was derived by Stehfest [56], who stated that
\begin{equation}
q(t) \approx \tilde{q}_{\Omega}(t) = \frac{\ln 2}{t} \sum_{n=1}^{\Omega} \kappa_{n} Q(n \frac{\ln 2}{t}), \quad \textrm{ where} 
\end{equation}
\begin{equation*}
\kappa_{n} = (-1)^{n+ \Omega/2} \sum_{k=[(n+1)/2]}^{\textrm{min}(n, \Omega/2)} \frac{k^{\Omega/2}(2k)!}{(\Omega/2 - k)!k!(n-k)!(2k-n)!},
\end{equation*}
and $\Omega$ is even. Note that the terms $\kappa_{n}$ can be extremely large, and that the value of $\kappa_{n}$ depends on the parameter $\Omega$. Furthermore, increasing the parameter $\Omega$ increases the accuracy of our estimate $q(t) \approx \tilde{q}_{\Omega}(t)$, provided that we have sufficient system precision to utilize the exact values for $\kappa_{n}$.

The Gaver-Stehfest algorithm is very efficient and accurate, but it requires high system precision for the weights $\kappa_{n}$ if it is to yield accurate estimates for $q(t)$. Indeed, if we wish to produce an estimate of $q(t)$ that is accurate to $N$ significant digits, we must calculate the values of $\kappa_{n}$ with an accuracy of about $2.5 N$ significant digits [1, 2]. Fortunately, to calculate $q(t)$ accurately we do not require such a disproportionately high level of accuracy in the values of $Q(s)$.

If the transform $Q(s)$ has all its singularities on the negative real axis, and if the function $q(t)$ is infinitely differentiable for all $t > 0$, extensive experimentation [1, 2] indicates that the relative error
\begin{equation}
\bigg| \frac{q(t) - \tilde{q}_{\Omega}(t)}{q(t)} \bigg| \approx 10^{-0.45 \Omega},
\label{error_from_GS}
\end{equation}
provided that the values $\kappa_{n}$ have been calculated with sufficient precision [1, 2]. If the function $q$ does not satisfy the above conditions $\tilde{q}_{\Omega}(t)$ may converge to $q(t)$ rather more slowly, but as a rule of thumb setting $\Omega = 10$ and using standard double precision for the weights $\kappa_{n}$ will ensure that the Gaver-Stehfest algorithm produces inversions that are accurate to at least three significant digits.

\section{Non-zero initial conditions}
\subsection{Non-zero initial conditions in a single edge} \label{non zero in an edge}
We now consider advection, diffusion and delivery along a single edge $ij$, where the initial condition $q_{ij}(x, 0)$ is non-zero. We let the length of $ij$ equal $l$, the longitudinal dispersion coefficient is $D$, the local delivery rate is $R$ and the mean velocity is $u$. 

We have seen that for any positive Laplace constant $s$, $Q_{1}(x,s) = e^{(g + h)\frac{x}{l} }$ and $Q_{2}(x,s) = e^{(g - h)\frac{x}{l} }$ satisfy the homogeneous analog, Equation (\ref{homogenous_laplace_of_CDC}). Furthermore, the Wronskian 
\begin{equation*}
W_{ij}(x,s) =  Q_{1}(x,s) \frac{\partial Q_{2}(x,s)}{\partial x} - \frac{\partial Q_{1}(x,s)}{\partial x} Q_{2}(x,s) = \frac{ - \alpha e^{2g \frac{x}{l} } }{D}.
\end{equation*}

By the method of variation of parameters, 
\begin{eqnarray}
f \big( x, s, q_{ij}(y,0) \big) 
& = & \frac{e^{(g - h) \frac{x}{l} }}{\alpha} \int_{0}^{x} e^{(h - g) \frac{y}{l} }  q_{ij}(y, 0) dy   \nonumber \\
&&- \frac{e^{(g + h) \frac{x}{l} }}{\alpha_{ij}} \int_{0}^{x} e^{-(g + h) \frac{y}{l} }  q_{ij}(y, 0) dy \qquad
\label{definition_of_f}
\end{eqnarray} 
is a particular solution to the fundamental Equation (\ref{laplace_of_CDC_equation}). 

Note that $f(0, s, q) = 0$ for all initial conditions $q$. Also note if $q = q_{1} + q_{2}$ then $f(x, s, q) = f(x, s, q_{1}) + f(x, s, q_{2})$. Since $f \big( x,s,q_{ij}(y,0) \big)$ is a particular solution of Equation (\ref{laplace_of_CDC_equation}), for each edge $ij$ there is a pair of constants $A$ and $B$ such that
\begin{equation}
Q_{ij}(x, s) =  A e^{(g + h) \frac{x}{l} } + B e^{(g - h) \frac{x}{l} } + f \big( x,s,q_{ij}(y,0) \big).
\label{definition_inhomo_Q}
\end{equation}

Because $f(0, s, q) = 0$ for all initial conditions $q$, Equations (\ref{definition_g_and_h}) and (\ref{definition_inhomo_Q}) imply that
\begin{equation}
X_{ij} \equiv Q_{ij}(0,s) = A + B, \qquad \textrm{and} 
\label{def_Ci_inhomog}
\end{equation}
\begin{eqnarray}
X_{ji} \equiv  Q_{ij}(l_{ij},s) & = & A e^{(g + h)} + B e^{(g - h)} +  f \big( l, s, q_{ij}(y,0) \big).
\label{def_Cj_inhomog}
\end{eqnarray}

We can therefore express $A$ and $B$ in terms of $X_{ij}$ and $X_{ji}$. Indeed, substituting Equation (\ref{def_Ci_inhomog}) into Equation (\ref{def_Cj_inhomog}) and multiplying both sides by $ e^{-g}$ tells us that
\begin{eqnarray}
X_{ji} e^{-g} & = & A \big(e^{h} - e^{-h} \big) + X_{ij} e^{-h} +  e^{-g} f \big( l, s, q_{ij}(y,0) \big).
\label{A_to_Qij_and_Qji}
\end{eqnarray}
We let
\begin{equation}
\beta_{ij}(s) \equiv \frac{- \alpha e^{-g}}{2  \textrm{ sinh} (h)} f \big( l, s, q_{ij}(y,0) \big),
\label{defn_beta}
\end{equation}
and explain its physical significance in Section \ref{Non zero initial conditions over a network}.

Equations (\ref{def_Ci_inhomog}) and (\ref{A_to_Qij_and_Qji}) imply that
\begin{eqnarray}
A & = & \frac{X_{ji}e^{-g} - X_{ij}e^{-h}}{2  \textrm{ sinh}(h)}  + \frac{\beta_{ij}}{\alpha} \quad \quad \textrm{ and} \nonumber \\
&& \nonumber \\
B & = & \frac{X_{ij} e^{h}  - X_{ji} e^{-g}}{2  \textrm{ sinh}(h)} - \frac{\beta_{ij}}{\alpha},
\label{def_A_and_B_inhomog}
\end{eqnarray}
and substituting Equation (\ref{def_A_and_B_inhomog}) into Equation (\ref{definition_inhomo_Q}) tells us that for any initial condition $q_{ij}(y, 0)$,
\begin{eqnarray}
Q_{ij}(x, s) & = &  \bigg( \frac{X_{ji}e^{-g} - X_{ij}e^{-h} }{2  \textrm{ sinh}(h)}  + \frac{\beta_{ij}}{\alpha} \bigg)  e^{(g + h) \frac{x}{l} }   \\
&& + \bigg( \frac{X_{ij} e^{h}  - X_{ji} e^{-g} }{2  \textrm{ sinh}(h)} - \frac{\beta_{ij}}{\alpha} \bigg) e^{(g - h) \frac{x}{l} }  +  f \big(x, s, q_{ij}(y,0) \big). \nonumber
\label{Qij}
\end{eqnarray}

\subsection{Non-zero initial conditions over a network} \label{Non zero initial conditions over a network}
Having analyzed the case of a single edge with a non-zero initial condition, we now consider an entire network, and find an exact solution that ensures that for all $t > 0$, the concentration varies continuously as we move from one edge to another. The first step in finding this solution is to note that Equation (\ref{definition_of_f}) implies that
 \begin{eqnarray}
 \frac{\partial  f \big(x, s, q_{ij} \big)}{\partial x} & = &
- \frac{(u + \alpha)}{2D \alpha} e^{(g + h) \frac{x}{l} } \int_{0}^{x}  e^{-(g + h) \frac{y}{l} } q_{ij}(y, 0) dy \nonumber \\
 && + \frac{(u - \alpha)}{2D \alpha} e^{(g - h) \frac{x}{l} } \int_{0}^{x}  e^{(h - g) \frac{y}{l} } q_{ij}(y, 0) dy, \nonumber
 \end{eqnarray}
where for the sake of clarity we drop the subscript $ij$ from $u_{ij}$, $\alpha_{ij}$, $l_{ij}$, $g_{ij}$, $h_{ij}$ and $D_{ij}$. Note that for any initial condition $q_{ij}(y, 0)$, we have $\frac{\partial  f \big(x, s, q_{ij}(y,0) \big)}{\partial x} \big|_{x=0} = 0$. It follows that
\begin{eqnarray}
\frac{\partial Q_{ij}(x,s)}{\partial x}\bigg|_{x=0}  
& = & \frac{\beta_{ij}}{D} + \frac{u +\alpha}{2D}  \bigg( \frac{X_{ji}e^{-g} - X_{ij}e^{-h} }{2  \textrm{ sinh}(h)} \bigg) \nonumber \\
&& +  \frac{u -\alpha}{2D} \bigg( \frac{X_{ij} e^{h}  - X_{ji} e^{-g} }{2  \textrm{ sinh}(h)}  \bigg).
\label{differentiate_Q_at_0}
\end{eqnarray}

Now, recall that $\Upsilon_{i}(s)$ denotes the Laplace transform of the net current of resource flowing away from node $i$, and that $\Upsilon_{i}(s) = 0$ unless $i$ is an inlet node. Substituting Equation (\ref{differentiate_Q_at_0}) into Equation (\ref{laplace_of_J}) gives us
\begin{eqnarray}
\Upsilon_{i}(s) 
& = & \sum_{j} X_{ij} \bigg[\frac{u_{ij}}{2} + \frac{\alpha_{ij} }{2  \textrm{ tanh}(h_{ij})} \bigg]   - \sum_{j} X_{ji} \frac{\alpha_{ij} e^{-g_{ij}} }{2  \textrm{ sinh}(h_{ij})}  - \sum_{j} \beta_{ij}(s). \qquad
\label{upsilon_in_terms_of_Qij}
\end{eqnarray}
Assuming that the cross-sectional areas $S_{ij}$ are constant, Equations (\ref{definition_Ci}) and (\ref{upsilon_in_terms_of_Qij}) imply that
\begin{eqnarray}
\Upsilon_{i}(s) & = & C_{i}(s) \sum_{j} S_{ij} \bigg(\frac{u_{ij}}{2} + \frac{\alpha_{ij} }{2  \textrm{ tanh}(h_{ij})} \bigg)   \nonumber \\
&& -  \sum_{j} C_{j}(s) S_{ij} \frac{\alpha_{ij} e^{-g_{ij}} }{2  \textrm{ sinh}(h_{ij})}  - \sum_{j} \beta_{ij}(s). \quad
\label{C_to_net_current}
\end{eqnarray}

In matrix form we have
\begin{equation}
\textbf{M}(s)\bar{C}(s) = \bar{p}(s), \qquad \textrm{where} 
\label{matrix_M_for_CDC}
\end{equation}
\begin{equation*}
\bar{C}(s) = \{C_{1}(s), C_{2}(s), \ldots , C_{m}(s) \}^{\textrm{T}}, 
\end{equation*}
\begin{equation}
p_{i}(s) = \Upsilon_{i}(s)  + \sum_{j} \beta_{ij}(s)
\label{defn_p}
\end{equation}
and $\textbf{M}(s)$ is the propagation matrix as in Equation (\ref{defn_M}). Note that the effect of the initial conditions on the concentration at the nodes is completely captured by the terms $\beta_{ij}(s)$, and that, as before, the propagation matrix $\textbf{M}(s)$ relates the concentrations at the nodes to the net currents flowing out of the nodes. Furthermore, by comparison with Equation (\ref{defn_M}), we can see that the concentration at the nodes is the same as would be the case if the network were initially empty, and the Laplace transform of the net current leaving node $i$ were $p_{i}(s)$ rather than $\Upsilon_{i}(s)$. 

In effect, the formalism of the propagation matrix enables us to substitute an initial condition in the edges around node $i$ for a boundary condition at node $i$. For each node $i$ and each Laplace variable $s$, this boundary condition is of the form $\sum_{j} \beta_{ij}(s)$. Intuitively speaking, the term $\beta_{ij}(s)$ represents the Laplace transform of the quantity of resource that first leaves edge $ij$ by arriving at node $i$. Note that we have not calculated the impact of the initial condition $q_{ij}(x,0)$ on the future concentration profile $q_{ij}(x,t)$ for $t>0$: we have simply calculated the impact of the initial conditions on the concentrations at the nodes (see Section \ref{efficient algorithm}). 

Since $\alpha_{ij}(s) \gg u_{ij}$ and $h_{ij}(s) \gg g_{ij}$ for large $s$, for very short time steps $t$ we have $\textrm{ sinh}(h_{ij}) \gg \max \big[ e^{g_{ij}}, e^{-g_{ij}} \big]$. It follows that over short time scales, the off-diagonal elements of $\textbf{M}(s)$ will be very small. If the entries in the $i$th column  of $\textbf{M}(s)$ are very small, it may be numerically difficult to calculate $C_{i}(s)$, as any error in our estimate for $C_{i}(s)$ would have very little impact on the value of $\textbf{M}(s) \bar{C}(s)$. 

In practice this is not a significant problem, as when we solve the above system of linear equations to identify $C_{i}(s)$, we make the initial guess that $C_{i}(s) = c_{i}(0)/s$, which would be the correct value if the concentration at node $i$ remained constant. For numerical reasons we may not be able to identify the exact value of $C_{i}(s)$, but this problem only arises when the bulk of resource around node $i$ does not leave the edges around node $i$ over the time scale $1/s$. As we shall see in Section \ref{leaves edge}, those are precisely the circumstances under which the value of $C_{i}(s)$ has little impact on our calculation of the spatial distribution of resource within the edges at a given time $t \approx 1/s$.

\section{Efficient calculation of resource distribution} \label{efficient algorithm}
If we wish to find the concentration at various points in the network other than the nodes, there are two ways we can proceed. The first method is to treat each point of interest as an additional node. The problem with this approach is that it increases the size of the propagation matrix, and finding $\bar{C}(s)$ by inverting the matrices $\textbf{M}(s)$ is the major computational cost of the propagation matrix algorithm. Furthermore, although this approach can be used to find the exact concentration at each of a given set of points, it does not provide a means of finding the exact quantity of resource between a given pair of points. We could approximate the total quantity of resource between two points by assuming that the concentration varies in a linear manner from one point to the next, but as the exact solution may contain boundary layer effects, we might require a very high spatial resolution to ensure that such a linear approximation is accurate.

A different approach, which we take, provides an exact solution for the total quantity of resource within each section of the network, regardless of the spatial resolution. The key conceptual step involves partitioning the resource into two parts. Strictly speaking our approach is mathematically continuous, but we can imagine that the resource is composed of particles, which either leave or do not leave a given edge over a given time scale. We let $\hat{q}_{ij}(x,t)$ denote the quantity of resource per unit length at the point $0 \leq x \leq l_{ij}$ in edge $ij$ and time $t$, where a given particle only contributes to $\hat{q}_{ij}(x,t)$ if it has passed through a node (any node) by time $t$ after initialization. More precisely, we work in Laplace space and let $\mathcal{L}\big(\hat{q}_{ij}(x,t) \big) = \hat{Q}_{ij}(x,s)$. This term denotes the Laplace transformed concentration profile that would occur if the network was initially empty, and if the Laplace transform of the net current leaving each node was $p_{i}(s) = \Upsilon_{i}(s)  + \sum_{j} \beta_{ij}(s)$, rather than $\Upsilon_{i}(s)$. 

As we have seen, the impact of the initial condition on the concentration at the nodes is completely captured by the constants $\beta_{ij}(s)$. However, $\hat{Q}_{ij}(x,s)$ and $\hat{q}_{ij}(x,t)$ do not fully capture the influence of the initial condition $q_{ij}(x,0)$ on the concentration profile $q_{ij}(x,t)$ for $t > 0$. In addition to $\hat{q}_{ij}(x,t)$ (the quantity of resource that has reached a node over the time scale $t$), we must also consider the resource that starts in edge $ij$, and which does not reach node  $i$ or $j$ over the time scale $t$. We let $\tilde{q}_{ij}(x,t)$ denote the quantity of such resource at the point $0 \leq x \leq l_{ij}$ in edge $ij$ and time $t$, where by definition \begin{equation}
 \tilde{q}_{ij}(x,t) = q_{ij}(x,t) - \hat{q}_{ij}(x,t). 
 \label{q as q_hat plus q_tilde}
\end{equation}

We can calculate the concentration at each node by calculating $\beta_{ij}(s)$ for every $i$ and $j$, and by using the propagation matrix. Furthermore, because at time 0 none of the resource in edge $ij$ has had time to reach a node, we can apply Equation \ref{Cij_as_a_function_of_x}, and find $\hat{Q}_{ij}(x,s)$ in terms of the boundary conditions $X_{ij}(s)$ and $X_{ji}(s)$. Given $\hat{Q}_{ij}(x,s)$ for $s = \frac{\ln 2}{t}, \ldots n \frac{\ln 2}{t}$, we can apply the Gaver-Stehfest algorithm and find $\hat{q}_{ij}(x,t)$. In addition, we solve a separate PDE for each edge, which tells us how the resource that stays within each edge has evolved over a given time step $t$. That is to say, for each edge $ij$ we find $\tilde{q}_{ij}(x,t)$, given that $\tilde{q}_{ij}(x,t)$ satisfies the fundamental advection-diffusion-delivery Equation (\ref{basic_CDE}), $\tilde{q}_{ij}(x,0) = q_{ij}(x,0)$, $\tilde{q}_{ij}(0,t) = 0$ and $\tilde{q}_{ij}(l_{ij}, t) = 0$. Finally, Equation \ref{q as q_hat plus q_tilde} tells us that  $q_{ij}(x,t) = \tilde{q}_{ij}(x,t) + \hat{q}_{ij}(x,t)$. 

In particular, we consider the case where the initial condition is stepwise constant, and edge $ij$ is divided into $N_{ij}$ sections of equal length. We let $k_{ij}^{(n)}(t)$ denote the mean quantity of resource per unit length in the $n$th section at the given time $t$, where by convention the first section is next to node $i$ and the $N$th section is next to node $j$. For any $t > 0$, we can employ the following algorithm to find an exact solution for the updated mean quantities per unit length, 
\begin{equation}
k_{ij}^{(n)}(t) = \frac{N_{ij}}{l_{ij}} \int_{\frac{n-1}{N_{ij}}l_{ij}}^{\frac{n}{N_{ij}}l_{ij}} q_{ij}(x, t) dx.
\label{defn_k_hat}
\end{equation}

\subsection{Stepwise constant initial conditions} \label{stepwise initial}
We are interested in calculating how the quantity of resource in a network changes over time, given that the resource decays and is subject to advection and diffusion. In particular, it is convenient to consider a stepwise constant initial condition, as we can then calculate how the total quantity of resource in each segment of the network has changed by time $t$. The first step in this calculation is to find the Laplace transform of the concentrations at each node $\bar{C}(s)$. As we have seen, to calculate $\bar{C}(s)$ we must first find $\textbf{M}_{ij}(s)$ and $\bar{\Upsilon}(s)$, which do not depend on the initial condition. For each sample point $s$ and each edge $ij$ we must also calculate $\beta_{ij}(s)$ and $\beta_{ji}(s)$, which capture the effect of the initial condition $q_{ij}(x,0)$. In particular, we start this subsection by considering the case where the initial condition is 
\begin{equation*}
q_{ij}(x,0) =  \Bigg\{	
\begin{array}{cl}
k & \textrm{if $\frac{n-1}{N} l_{ij} \leq x < \frac{n}{N} l_{ij}$} \\
0 & \textrm {otherwise,}
\end{array}
\end{equation*} 
where $n \leq N$. We will find our solutions for other initial conditions by summing the solutions for various initial conditions of this form. For the sake of clarity we drop the subscripts $ij$ from $l_{ij}$, $N_{ij}$, $g_{ij}$ and $h_{ij}$, and note that Equation (\ref{definition_of_f}) tells us that for this initial condition
\begin{eqnarray}
f \big( l, s, q_{ij} \big) & = & -\frac{k e^{g + h}}{\alpha_{ij}} \int_{\frac{n-1}{N} l}^{\frac{n}{N} l} e^{-(g + h) \frac{x}{l} } dx  +  \frac{k e^{g - h}}{\alpha_{ij}} \int_{\frac{n-1}{N} l}^{\frac{n}{N} l} e^{(h - g) \frac{x}{l} } dx, \nonumber \\
& = & \frac{2 D_{ij} k e^{g + h}}{\alpha_{ij} (u_{ij} + \alpha_{ij})} \bigg( e^{\frac{-n}{N}(g + h) } - e^{\frac{-(n-1)}{N}(g + h)} \bigg) \nonumber \\
&& - \frac{2 D_{ij} k e^{g + h}}{\alpha_{ij} (u_{ij} - \alpha_{ij})} \bigg( e^{\frac{n}{N}(h - g) } - e^{\frac{(n-1)}{N}(h - g) } \bigg). \nonumber \\
\label{g_constant_initial_condition}
\end{eqnarray}

Substituting into Equation (\ref{defn_beta}) yields
\begin{eqnarray}
\beta_{ij}(s) & = & \frac{ k e^{\frac{1-n}{N} g}}{4 (s + R_{ij}) \textrm{ sinh}(h)} \times  \\
&& \bigg[ e^{\frac{N-n}{N} h} \big(e^{\frac{h}{N}} - e^{\frac{-g}{N}}  \big) \big(\alpha_{ij} - u_{ij} \big)  + \quad e^{\frac{n-N}{N} h} \big(e^{\frac{-h}{N}} - e^{\frac{-g}{N}}  \big) \big(\alpha_{ij} + u_{ij} \big) \bigg]. \nonumber
\label{beta_step_constant_initial_condition}
\end{eqnarray}
\\
Recall that $f(x, s, q_{1} + q_{2}) = f(x, s, q_{1}) + f(x, s, q_{2})$. Since Equation (\ref{defn_beta}) is linear, it follows that if the initial condition contains several blocks of resource, each block makes its own separate contribution to $\beta_{ij}(s)$ and $\beta_{ji}(s)$. Let $x_{0} = 0, x_{1} = \frac{l}{N}, x_{2} = \frac{2l}{N}, \ldots , x_{N} = l$, and suppose that for all $1 \leq n \leq N$ we have
\begin{equation}
q_{ij}(x, 0) = k_{ij}^{(n)} \quad \textrm{ for all} \quad  x_{n-1} < x < x_{n}. 
\label{stepwise}
\end{equation}

Given such a stepwise constant initial condition, we can calculate $\beta_{ij}(s)$ by summing the contribution of each of the blocks of resource. That is to say, Equation (\ref{beta_step_constant_initial_condition}) becomes
\begin{eqnarray}
\beta_{ij}(s) & = & \sum_{n = 1}^{N} \frac{ k_{ij}^{(n)} e^{\frac{1-n}{N}g_{ij}}}{4 (s + R_{ij}) \textrm{ sinh}(h_{ij})}  \bigg[ e^{\frac{N-n}{N} h_{ij}} \big(e^{\frac{h_{ij}}{N}} - e^{\frac{-g_{ij}}{N}}  \big) \big(\alpha_{ij} - u_{ij} \big) \nonumber \\
&& \qquad + \quad e^{\frac{n-N}{N} h_{ij}} \big(e^{\frac{-h_{ij}}{N}} - e^{\frac{-g_{ij}}{N}}  \big) \big(\alpha_{ij} + u_{ij} \big) \bigg]. 
\label{beta_ij}
\end{eqnarray}

We can find $\beta_{ji}(s)$ by using the above formula, substituting $-g_{ij}$ for $g_{ji}$, $-u_{ij}$ for $u_{ji}$ and $k_{ij}^{(N - n + 1)}$ for $k_{ji}^{(n)}$. It follows that
\begin{eqnarray}
\beta_{ji}(s) & = & \sum_{n = 1}^{N} \frac{ k_{ij}^{(N - n + 1)} e^{\frac{n-1}{N} g_{ij}}}{4 (s + R_{ij}) \textrm{ sinh}(h_{ij})} \bigg[ e^{\frac{N-n}{N} h_{ij}} \big(e^{\frac{h_{ij}}{N}} - e^{\frac{g_{ij}}{N}}  \big) \big(\alpha_{ij} + u_{ij} \big) \nonumber \\
&& \qquad + \quad e^{\frac{n-N}{N} h_{ij}} \big(e^{\frac{-h_{ij}}{N}} - e^{\frac{g_{ij}}{N}}  \big) \big(\alpha_{ij} - u_{ij} \big) \bigg]. 
\label{beta_ji}
\end{eqnarray}

\subsection{Resource that leaves its initial edge} \label{leaves edge}
If a particle leaves edge $ij$ and reaches node $i$ or $j$ over the relevant time scale, it contributes to $\beta_{ij}(s)$ or $\beta_{ji}(s)$,  and hence it contributes to our solution $C_{i}(s)$, $C_{j}(s)$ and $\mathcal{L}\big(\hat{q}_{ij}(x,t) \big) = \hat{Q}_{ij}(x,s)$. On the other hand, at time 0 none of the resource has reached the nodes, so the initial condition $\hat{q}_{ij}(x, 0) = 0$. It follows that if the cross-sectional areas are held constant, we can apply Equation (\ref{Cij_as_a_function_of_x}). In other words, we can find $\hat{Q}_{ij}(x,s)$ by effectively considering an initially empty network, where resource is introduced at the nodes at a rate which exactly matches the rate at which resource reaches the nodes in the case where the network has the given non-zero initial condition. Equation (\ref{Cij_as_a_function_of_x}) also accounts for the impact of any inlet nodes, in the case where resource is being added to the network. 

We can therefore use Equations (\ref{matrix_M_for_CDC}), (\ref{beta_ij}) and (\ref{beta_ji}) to find $\bar{C}(s) = \{ C_{1}(s), \ldots , C_{m}(s) \}$, and in the case where the cross-sectional areas are constant, we can use Equations (\ref{Cij_as_a_function_of_x}) and (\ref{definition_Ci}) to express $\hat{Q}_{ij}(x,s)$ in terms of the boundary conditions $X_{ij} = S_{ij} C_{i}(s)$ and $X_{ji} = S_{ij} C_{j}(s)$. Since $\mathcal{L}\big( \int \hat{q}_{ij}(x, t) dx \big) =  \int \hat{Q}_{ij}(x,s) dx$, we can find $\int \hat{q}_{ij}(x, t) dx$ by calculating $\int \hat{Q}_{ij}(x, s) dx$ for $s = \ln 2/t, \ldots , N \ln 2/t$ and applying the Gaver-Stehfest algorithm. 

We suppose that edge $ij$ is divided into $N_{ij}$ sections of equal length, and for the sake of clarity we drop the subscripts $ij$ from $D_{ij}$, $l_{ij}$ and $N_{ij}$. We let $y_{ij}^{(n)}(t)$ denote the mean value of $\hat{q}_{ij}(x,t)$ in the $n$th section of edge $ij$, and note that by definition 
\begin{equation}
y_{ij}^{(n)}(t) = \frac{N}{l}  \int_{\frac{n-1}{N}l}^{\frac{n}{N}l} \hat{q}_{ij}(x,t) dx.
\end{equation}
Defining $Y_{ij}^{(n)}(s) \equiv  \mathcal{L} \big( y_{ij}^{(n)}(t) \big)$ we have
\begin{eqnarray}
Y_{ij}^{(n)}(s) & = & \frac{N}{l}  \int_{\frac{n-1}{N}l}^{\frac{n}{N} l} \hat{Q}_{ij}(x,s) dx \nonumber \\
& = & \frac{N X_{ij}}{l \textrm{ sinh}(h_{ij}) } \int_{\frac{n-1}{N} l}^{\frac{n}{N} l}  \textrm{ sinh}\big(h_{ij} \frac{l - x}{l} \big) e^{g_{ij} \frac{x}{l} } dx \nonumber \\
&&   + \quad  \frac{N X_{ji}}{l \textrm{ sinh}(h_{ij}) } \int_{\frac{n-1}{N} l}^{\frac{n}{N} l} \textrm{ sinh}\big(h_{ij} \frac{x}{l} \big) e^{g_{ij} \frac{x - l}{l} } dx \nonumber \\
 && \nonumber \\
& = & \frac{N D}{l \textrm{ sinh}(h_{ij}) } \bigg[ \frac{X_{ij} e^{h_{ij}} - X_{ji}   e^{-g_{ij}} }{u_{ij} - \alpha_{ij} } e^{(g_{ij} - h_{ij}) \frac{x}{l} } \nonumber \\
 && \qquad + \quad \frac{X_{ji} e^{-g_{ij}} - X_{ij} e^{-h_{ij}} }{u_{ij} + \alpha_{ij} } e^{(g_{ij} + h_{ij}) \frac{x}{l} } \bigg]_{\frac{n-1}{N} l}^{\frac{n}{N} l}, \nonumber
 \end{eqnarray}
which implies that
 \begin{eqnarray}
 Y_{ij}^{(n)}(s)  & = &  \eta_{ij}(s) \big( \alpha_{ij} + u_{ij} \big) \times \nonumber \\
  && \bigg[ X_{ij} \big( e^{\frac{n-1}{N}(g_{ij} - h_{ij})} - e^{\frac{n}{N}(g_{ij} - h_{ij})} \big)  + X_{ji} \times \nonumber \\
  && \big( e^{\frac{n-N}{N}g_{ij} - \frac{n+N}{N}h_{ij}} - e^{\frac{n-N-1}{N}g_{ij} - \frac{n+N-1}{N}h_{ij}} \big) \bigg]\nonumber \\
  & + & \eta_{ij}(s) \big( \alpha_{ij} - u_{ij} \big) \times \nonumber \\
 && \bigg[ X_{ij} \big( e^{\frac{n - 1}{N} g_{ij} - \frac{2N-n+1}{N}h_{ij}} - e^{\frac{n}{N}g_{ij}  - \frac{2N - n}{N}h_{ij}} \big) \nonumber \\ 
& + & X_{ji} \big( e^{\frac{n-N}{N}(g_{ij} + h_{ij})} - e^{\frac{n-N-1}{N} (g_{ij}+h_{ij})} \big) \bigg],
\label{int_Q}
\end{eqnarray}
\begin{equation}
\textrm{where} \quad \eta_{ij}(s) = \frac{N_{ij} e^{h_{ij}} }{4 (s + R_{ij}) l_{ij} \textrm{ sinh}\big( h_{ij} \big) }.
\label{defn_eta}
\end{equation} 

\subsection{Resource that remains in its initial edge} \label{stays in edge}
Over the time scale $t$, not all of the resource will leave the edge in which it started. To find $\tilde{q}_{ij}(x,t)$, the quantity of resource that has not left edge $ij$, we must solve the advection, diffusion, delivery problem for each separate edge $ij$, where nodes $i$ and $j$ are absorbing boundaries and the initial condition $\tilde{q}_{ij}(x, 0) = q_{ij}(x, 0)$. The resulting solution accounts for those particles which do not reach a node in the relevant time-scale. In particular, we consider the case where the initial condition is stepwise constant, as in Equation (\ref{stepwise}). 

The fundamental Equation (\ref{basic_CDE}) tells us that for each edge
\begin{equation}
\frac{\partial}{\partial t} \tilde{q}_{ij} = D_{ij} \frac{\partial^{2}}{\partial x^{2}} \tilde{q}_{ij} - u_{ij} \frac{\partial}{\partial x}\tilde{q}_{ij} - R_{ij} \tilde{q}_{ij}.
\label{fundamental_equation_for_q}
\end{equation}
Furthermore, we are looking for a real valued function such that $\tilde{q}_{ij}(0,t) = 0$ and $\tilde{q}_{ij}(l_{ij},t) = 0$ for all $t$. These conditions imply that we can express $\tilde{q}_{ij}(x, t)$ in the following form:
\begin{eqnarray}
\tilde{q}_{ij} (x,t) & = & e^{\frac{u_{ij}}{2D_{ij}}  x } \sum_{m = 1}^{\infty} A^{m} e^{\lambda_{ij}^{m} t} \textrm{sin}\big(\frac{m \pi x}{l_{ij}} \big), \nonumber \\
\textrm{where} \quad \lambda_{ij}^{m} & = & - \bigg( m^{2} \frac{D_{ij} \pi^{2} }{ l_{ij}^{2} } + \frac{ u_{ij}^{2} }{4D_{ij}} + R_{ij} \bigg).
\label{defn_tilde_q}
\end{eqnarray}

The parameters $A^{m}$ can be found by taking Fourier transforms. More specifically, we know that $\tilde{q}_{ij}(x, 0) = q_{ij}(x, 0)$, so
\begin{eqnarray*}
\sum_{n = 1}^{\infty} A^{m} \textrm{sin}\big(\frac{m \pi x}{l_{ij}} \big) & = & q_{ij}(x, 0) e^{-g_{ij} \frac{x}{l_{ij}} } \quad \textrm{ and} \\
\int_{0}^{l} \textrm{sin}\big(\frac{m \pi x}{l_{ij}}\big) \textrm{ sin}\big(\frac{n \pi x}{l_{ij}}\big) dx & = & \Bigg\{	
\begin{array}{cl}
0 & \textrm{ if $m \neq n$,} \\
& \\
\frac{l_{ij}}{2} & \textrm{ if $m = n$.}  
\end{array}
\end{eqnarray*}

It follows that for every positive integer $m$,
\begin{equation*}
A^{m} = \frac{2}{l_{ij}} \int_{0}^{l_{ij}} \textrm{ sin}\big(\frac{m \pi x}{l_{ij}}\big) q_{ij}(x, 0) e^{-g_{ij} \frac{x}{l_{ij}} } dx. \end{equation*}
In particular, consider the case where the initial condition is stepwise constant, and of the form described by Equation (\ref{stepwise}). We have
\begin{eqnarray}
A^{m} & = & \mu_{ij}^{m} \sum_{n = 1}^{N_{ij}} k_{ij}^{(n)} \bigg[ e^{-g_{ij} \frac{x}{l_{ij}} }  \bigg( \frac{-g_{ij}}{\pi m} \textrm{sin} \big(\frac{m \pi x}{l_{ij}} \big) - \textrm{cos} \big(\frac{m \pi x}{l_{ij}} \big) \bigg) \bigg]_{\frac{n-1}{N_{ij}}l_{ij}}^{\frac{n}{N_{ij}}l_{ij}} \nonumber \\
&& \nonumber \\
& = & \mu_{ij}^{m} \sum_{n = 1}^{N_{ij}-1} \bigg[ e^{\frac{-n}{N_{ij}} g_{ij} } \bigg( k_{ij}^{(n+1)} - k_{ij}^{(n)} \bigg)  \bigg( \frac{g_{ij}}{\pi m} \textrm{sin} \big(\frac{m n \pi}{N_{ij}} \big) + \textrm{cos} \big(\frac{m n \pi }{N_{ij}} \big) \bigg) \bigg] \nonumber  \\
&& + \quad \mu_{ij}^{m} \bigg( k_{ij}^{(1)} -  k_{ij}^{(N_{ij})} e^{-g_{ij} } (-1)^{m} \bigg), 
 \label{defn_An}
\end{eqnarray}
where
\begin{equation}
\mu_{ij}^{m} =   \frac{8 D_{ij}^{2} \pi m}{u_{ij}^{2} l_{ij}^{2} + 4 D_{ij}^{2} \pi^{2} m^{2}}.
\label{defn_mu}
\end{equation} 

We are now in a position to find 
\begin{equation*}
z_{ij}^{(n)}(t) = \frac{N_{ij}}{l_{ij}}  \int_{\frac{n-1}{N_{ij}}l_{ij}}^{\frac{n}{N_{ij}}l_{ij}} \tilde{q}_{ij}(x,t) dx,
\end{equation*} 
as Equation (\ref{defn_tilde_q}) implies that
\begin{eqnarray}
z_{ij}^{(n)}(t) & = & \frac{N_{ij}}{l_{ij}} \int_{\frac{n-1}{N_{ij}}l_{ij}}^{\frac{n}{N_{ij}}l_{ij}} e^{\frac{g_{ij} x}{l_{ij}} } \sum_{m = 1}^{\infty} A^{m} e^{\lambda_{ij}^{m} t} \textrm{sin}\big(\frac{m \pi x}{l_{ij}} \big)dx \nonumber \\
& = &  \frac{N_{ij}}{2} e^{g_{ij} \frac{n}{N_{ij}}} \sum_{m = 1}^{\infty} \mu_{ij}^{m} A^{m} e^{\lambda_{ij}^{m} t} \bigg[ \frac{g_{ij}}{\pi m} \times \nonumber \\
&&  \bigg( \textrm{sin}\big(\frac{m n \pi}{N_{ij}} \big) -  e^{\frac{-g_{ij}}{N_{ij}}} \textrm{sin}\big(\frac{m (n-1) \pi}{N_{ij}} \big) \bigg) \nonumber \\
&& + \bigg( e^{\frac{-g_{ij}}{N_{ij}}} \textrm{cos}\big(\frac{m (n-1) \pi}{N_{ij}} \big)- \textrm{cos}\big(\frac{m n \pi}{N_{ij}} \big) \bigg) \bigg]. \qquad
\label{k_int}
\end{eqnarray}

Note that $\mu_{ij}^{m} \rightarrow \frac{2}{\pi m}$ as $m \rightarrow \infty$, and likewise $A^{m} \in O(m^{-1})$. In contrast  $e^{\lambda_{ij}^{m} t}$ tends to zero much more rapidly. Indeed, we note that 
\begin{eqnarray}
\sum_{m = \Omega' }^{\infty} e^{\lambda_{ij}^{m} t} & = & e^{- \big( \frac{ u_{ij}^{2} }{4D_{ij}} + R_{ij} \big) t} \sum_{m = \Omega'}^{\infty} e^{- \frac{D_{ij}  \pi^{2} t}{ l_{ij}^{2} } m^{2} } \nonumber \\
& < & \frac{e^{- \big( \frac{ u_{ij}^{2} }{4D_{ij}} + R_{ij} \big) t} }{\Omega'} \int_{\Omega'}^{\infty} x e^{- \frac{D_{ij}  \pi^{2} t}{ l_{ij}^{2} } x^{2} } dx \nonumber \\
& < & \frac{l_{ij}^{2}  }{ 2 \Omega' \pi^{2} D_{ij} t } e^{ \lambda_{ij}^{\Omega'} t}.
\end{eqnarray}
It follows that the relative error
\begin{equation*}
\bigg| \frac{ \sum_{m = 1}^{\infty} e^{\lambda_{ij}^{m} t} - \sum_{m = 1}^{\Omega'} e^{\lambda_{ij}^{m} t} }{ \sum_{m = 1}^{\infty} e^{\lambda_{ij}^{m} t}} \bigg| <  \frac{ \sum_{m = \Omega'}^{\infty} e^{\lambda_{ij}^{m} t}  }{ \sum_{m = 1}^{\infty} e^{\lambda_{ij}^{m} t}} < \epsilon
\end{equation*}
whenever we have
\begin{equation}
e^{\lambda_{ij}^{\Omega'} t} <  \epsilon \frac{ 2 \Omega' \pi^{2} D_{ij} t }{ l_{ij}^{2}  } \sum_{m = 1}^{\Omega'} e^{\lambda_{ij}^{m} t}.
\label{truncation_error}
\end{equation}

We can therefore be confident that if we truncate the sum in Equation (\ref{k_int}) at $m = \Omega'$, the relative errors in our estimates for $z_{ij}^{(n)}(t)$ will be smaller than $\epsilon$ provided that $\Omega'$ satisfies Equation (\ref{truncation_error}). Also note that Equation (\ref{defn_tilde_q}) tells us that if $D_{ij} t > l_{ij}^{2}$ then $e^{\lambda_{ij}^{m} t}$ decreases rapidly, so $\Omega'$ does not need to be large unless $D_{ij} t \ll l_{ij}^{2}$. Furthermore, if $u_{ij}^{2} t^{2} > l_{ij}^{2}$ then most of the resource will leave edge $ij$ over the time scale $t$, and $\tilde{q}_{ij}(x,t)$ will only make a small contribution to the total value of $q_{ij}(x,t)$. In that case using a small value of $\Omega'$ will produce very accurate estimates for $k_{ij}^{(n)}(t)$ even if $D_{ij} t \ll l_{ij}^{2}$.

\subsection{Calculating the total quantity of resource in each segment of a network} 
Suppose that we wish to calculate the mean concentration per unit length in each segment of a network at time $t$, such that each part of our final answer has a relative error $\epsilon < 10^{-0.45 \Omega}$, where $\Omega$ is an even integer. The first step is to set $s = \Omega \ln 2/t$, and apply Equations (\ref{beta_ij}) and (\ref{beta_ji}) to find $\beta_{ij}(s)$ and $\beta_{ji}(s)$ for each edge $ij$. We then compute $\mathbf{M}(s)$ and $\bar{p}(s)$, and employ the BiCGStab algorithm to find $\bar{C}(s_{\Omega})$, starting with the initial guess that for each $i$,
\begin{equation}
C_{i}(s_{\Omega}) \approx \frac{\tau }{\Omega \ln 2 } c_{i}(0) = \frac{\tau }{\Omega \ln 2 } \frac{ \sum_{j} k_{ij}^{(1)} }{ \sum_{j} S_{ij}(0)}.
\label{guess_C_from_constant_c}
\end{equation}
This initial guess for the value of $\bar{C}(s_{\Omega})$ would be correct if the concentration at the nodes was constant, and making such a guess can help to speed up the process of finding the true value of $\bar{C}(s_{\Omega})$. At each step, when we have identified $\bar{C}(s)$ such that $\mathbf{M}(s)\bar{C}(s) = \bar{p}(s)$, we store the vector $\bar{C}(s)$ and repeat for $s = s_{\Omega-1}, \ldots , s_{1}$, where $s_{n} = n \ln 2/t$. The only difference is that for subsequent applications of the BiCGStab algorithm, we can take advantage of the approximation
\begin{equation}
C_{i}(s_{n}) \approx \frac{n+1}{n} C_{i}(s_{n+1}).
\end{equation}
This is generally a better initial guess than that provided by Equation (\ref{guess_C_from_constant_c}), so the BiCGStab algorithm converges on the solution more rapidly. Given $C_{i}(s_{n})$ and $C_{j}(s_{n})$, we can use Equation (\ref{int_Q}) to calculate $Y_{ij}^{(m)}(s_{n})$ for each section in the edge $ij$. Having found $Y_{ij}^{(m)}(s_{n})$  for each $1 \leq n \leq \Omega$, we can apply the Gaver-Stehfest algorithm to obtain $y_{ij}^{(m)}(t)$, and we repeat this process for each edge in the network. Finally, for each edge $ij$ we can use Equations (\ref{defn_tilde_q}), (\ref{defn_mu}) and (\ref{defn_An}) to calculate a sequence of values for $e^{\lambda_{ij}^{m} t}$, $\mu_{ij}^{m}$ and $A^{m}$ until we reach an integer $\Omega'$ such that $e^{\lambda_{ij}^{\Omega'} t}$ satisfies Equation (\ref{truncation_error}). We then employ Equation (\ref{k_int}) to find $z_{ij}^{(1)}(t), \ldots , z_{ij}^{(N_{ij})}(t)$  (the mean quantity of resource in $ij$ that has not reached a node), and note that for each section of the network the mean quantity of resource per unit length
\begin{equation}
k_{ij}^{(n)}(t) =  y_{ij}^{(n)}(t) + z_{ij}^{(n)}(t).
\end{equation}

Unless there are many sections in each edge, finding the vectors $\bar{C}(s)$ such that $\mathbf{M}(s)\bar{C}(s) = \bar{p}(s)$ is the most time consuming step of the computation, as it effectively involves inverting an $m \times m$ matrix $\mathbf{M}(s)$, where $m$ is the number of nodes. We also note that Equation (\ref{defn_M}) implies that if $h_{ij}$  is larger than 10 (say), then the matrix $\mathbf{M}(s)$ may be close to singular, making it  computationally difficult to calculate $\bar{C}(s)$. Fortunately this problem is easy to avoid, as we can simply introduce an additional node $k$ at the midpoint of edge $ij$. This increases the size of the matrix $\mathbf{M}(s)$, but the lengths $l_{ik}$ and $l_{kj}$ will be half the length $l_{ij}$. As we have seen, the ratio $\mathbf{M}_{ij}(s):\mathbf{M}_{ji}(s)$ is equal to $1:e^{2g_{ij}}$, so adding additional nodes greatly reduces the ratio between the entries of $\mathbf{M}(s)$, and can make it significantly easier to find the vector $\bar{C}(s)$.

Finally, we note that this algorithm can be adapted for the case where the cross-sectional areas $S_{ij}(t)$ vary continuously over time (see Section IV). However, even in the case where $S_{ij}(t)$ varies continuously over time, our method requires that over each time step the lengths $l_{ij}$, mean velocities $u_{ij}$, decay rates $R_{ij}$ and dispersion coefficients $D_{ij}$ are held constant. In the case where we wish to find the concentration of resource in a changing network, we simply vary all the parameters in a stepwise manner, finding the spatial distribution of resource at the end of each time step, and treating that distribution as an initial condition for the following time step. In the case of the fungal networks that we analyze in the Main Text, this approach yields very similar results to the more complex algorithm with continuously varying $S_{ij}(t)$ that we outline in the following section.

\section{Advection, diffusion and delivery in a changing network.} \label{changing network}
We now consider the case where each cross-sectional area $S_{ij}(t)$ varies monotonically over time, while the lengths $l_{ij}$, mean velocities $u_{ij}$, decay rates $R_{ij}$ and dispersion coefficients $D_{ij}$ remain constant. Equation (1) tells us that the dispersion coefficients $D_{ij}$ will be constant and equal to the molecular diffusion coefficient $D_{m}$ if the edges are sufficiently narrow or the velocities sufficiently low. Alternatively, $D_{ij}$ and $u_{ij}$ would remain constant despite the changing cross-sections in the specific, but biologically relevant case where the edges are composed of some variable number of tubes of fixed radius $r_{ij}$, while the pressure at each node remains constant over time [26, 27, 44]. In that case the conductance of $ij$ will be proportional to $S_{ij}(t)$, so the medium-current at time $t$ will be proportional to the pressure drop times $S_{ij}(t)$, and the velocity $u_{ij}$ will be constant.

More generally, if we are considering advection, diffusion and delivery over a network where the parameters $u_{ij}$, $S_{ij}$, $R_{ij}$ and $D_{ij}$ change over time, it is reasonable to assume that $u_{ij}$, $R_{ij}$ and $D_{ij}$ are piece-wise constant provided that the time scales for transiting the edges $ij$ are small compared to the time scales over which $u_{ij}$, $R_{ij}$ and $D_{ij}$ are changing. For example, in the case of vascular networks the cross-sectional areas of capillaries, the velocity of blood flow and the rates of resource delivery may vary over time, but such changes occur over time scales that are large compared to the time it takes to transit a capillary. In such a case $R_{ij}$ might represent the local rate of glucose delivery per unit of glucose in the blood (for example), and the following algorithm enables us to calculate the concentrations that arise at times $t_{1}$, $t_{2}$, etc, as we vary $S_{ij}(t)$ in a continuous manner, while $u_{ij}$, $R_{ij}$ and $D_{ij}$ vary in a stepwise manner, being held constant between each of the time points of interest.

Now, suppose that we want to know how the spatial distribution of resource in a network changes over a time-scale $\tau$. We let $S_{ij}(t)$ denote the area of edge $ij$ at time $t$, and where $S_{ij}(0)$ and $S_{ij}(\tau)$ are given quantities, it is mathematically convenient to set
\begin{eqnarray}
S_{ij}(t) & = & \big( 2S_{ij}(\tau) - S_{ij}(0) \big) + \big(2S_{ij}(0) - 2S_{ij}(\tau) \big) e^{\frac{- \ln 2}{\tau}t} \nonumber \\
& = & a_{ij} + b_{ij}e^{\frac{- \ln 2}{\tau}t}.
\label{defn_St}
\end{eqnarray}

By adopting this functional form for $S_{ij}(t)$, we are assuming that the cross-sectional areas $S_{ij}$ vary in an approximately linear manner over the time scale of interest $\tau$. In fact, the rate of change $\frac{d}{d t}S_{ij}$ halves over the time scale $0 \leq t \leq \tau$. Also note that if the given cross-sectional areas $S_{ij}(0)$ and $S_{ij}(\tau)$ are non-negative, it follows that $S_{ij}(t) \geq 0$ for all $0 \leq t \leq \tau$. However, as $t \rightarrow \infty$, $S_{ij}(t) \rightarrow 2S_{ij}(\tau) - S_{ij}(0)$, which may be negative. Since the value of $S_{ij}(t)$ has little effect on our calculations for $t > \tau$, we are not introducing a major source of error when we allow the possibility of negative values for  $S_{ij}(t)$ at time points beyond the time of interest. 
 
 \subsection{Propagation matrices for a changing network} \label{Tranform for changing areas}
Suppose that the cross-sectional areas $S_{ij}(t)$ are of the form described by Equation (\ref{defn_St}), and $u_{ij}$ and $D_{ij}$ are constant. By definition we have $q_{ij}(x,t) = S_{ij}(t)c_{ij}(x,t)$, so taking Laplace transforms gives us
\begin{eqnarray}
Q_{ij}(x, s) & = & \mathcal{L}\big( a_{ij} c_{ij}(x,t) \big) + \mathcal{L}\big( b_{ij} e^{\frac{- \ln 2}{\tau}t}c_{ij}(x,t) \big) \nonumber \\
&& \nonumber \\
& = & a_{ij} C_{ij} \big(x,s \big) + b_{ij} C_{ij} \big(x, s + \ln 2/\tau \big).
\label{definition_Q_changing_S}
\end{eqnarray}
In particular, writing $s' = s + \ln 2/\tau$, we define
\begin{equation}
X_{ij}(s) \equiv Q_{ij}(0, s) = a_{ij} C_{i}(s) + b_{ij} C_{i}(s') \qquad \textrm{ and} \\
\label{defn_Qij_changing_S}
\end{equation}
\begin{equation}
X_{ji}(s) \equiv Q_{ij}(l_{ij}, s) = a_{ij} C_{j}(s) + b_{ij} C_{j}(s').
\label{defn_Qji_changing_S}
\end{equation}

Substituting Equations (\ref{defn_Qij_changing_S}) and (\ref{defn_Qji_changing_S}) into Equation (39) tells us that
\begin{eqnarray}
\Upsilon_{i}(s) & = & \sum_{j} \bigg( a_{ij} C_{i}(s) + b_{ij} C_{i}(s') \bigg) \bigg(\frac{u_{ij}}{2} + \frac{\alpha_{ij} }{2  \textrm{ tanh}\big( h_{ij} \big)} \bigg)   \nonumber \\
&& -  \quad \sum_{j} \bigg( a_{ij} C_{j}(s) + b_{ij} C_{j}(s') \bigg) \frac{\alpha_{ij} e^{-g_{ij}} }{2  \textrm{ sinh}\big(h_{ij} \big)}  -  \sum_{j} \beta_{ij}(s). \qquad
\end{eqnarray}

In matrix form we have the equivalent of Equation (41):
\begin{equation}
\textbf{V}(s)\bar{C}(s) + \textbf{W}(s)\bar{C}(s + \ln 2/\tau) = \bar{p}(s),  
\label{matrix_N_for_CDC_vary}
\end{equation}
\begin{equation*}
\textrm{where} \quad \bar{C}(s) = \{C_{1}(s), C_{2}(s), \ldots , C_{m}(s) \}^{\textrm{T}}, 
\end{equation*}
\begin{equation*}
p_{i}(s) = \Upsilon_{i}(s)  + \sum_{j} \beta_{ij}(s),
\end{equation*}
\begin{equation}
\textbf{V}_{ij}(s) =  \Bigg\{	
\begin{array}{cl}
\sum_{k} a_{ij} \bigg[ \frac{u_{ik}}{2} + \frac{\alpha_{ik}}{2  \textrm{ tanh} ( h_{ik} )} \bigg] & \textrm{ if $i = j$,} \\
& \\
\frac{- a_{ij} \alpha_{ij} e^{-g_{ij}}}{2  \textrm{ sinh} \big( h_{ij} \big)} & \textrm{ otherwise,}  
\end{array}
\label{defn_V}
\end{equation}
and
\begin{equation}
\textbf{W}_{ij}(s) =  \Bigg\{	
\begin{array}{cl}
\sum_{k} b_{ij} \bigg[ \frac{u_{ik}}{2} + \frac{\alpha_{ik}}{2  \textrm{ tanh} ( h_{ik} )} \bigg] & \textrm{ if $i = j$,} \\
& \\
\frac{- b_{ij} \alpha_{ij} e^{-g_{ij}}}{2  \textrm{ sinh} \big(h_{ij} \big)} & \textrm{ otherwise.}  
\end{array}
\label{defn_W}
\end{equation}

\subsection{Calculating $\bar{C}(s)$ in a changing network} \label{find C in changing network}
To find the boundary conditions $X_{ij}(s)$ and $X_{ji}(s)$ for each edge, we must first calculate the Laplace transform of the concentration at each node. Furthermore, in order to apply the Gaver-Stehfest algorithm, we need to calculate $\bar{C}(s)$ for $s = s_{1}, \ldots , s_{\Omega}$ where $s_{n} = n \ln 2 /\tau$. As in the case where the cross-sectional areas of the network remain constant, we can calculate $\textbf{V}(s)$, $\textbf{W}(s)$ and $\bar{p}(s)$ for any positive integer $s$. Since $s_{n+1} =  s_{n} + \ln 2 /\tau$, we can use Equation (\ref{matrix_N_for_CDC_vary}) to relate the vectors $\bar{C}(s_{n+1})$ and $\bar{C}(s_{n})$. That is to say, for each $n$ we have
\begin{equation}
\textbf{V}(s_{n})\bar{C}(s_{n}) = \bar{p}(s_{n}) - \textbf{W}(s_{n})\bar{C}(s_{n+1}).  
\label{iterative_C}
\end{equation}
To begin this iterative process of finding $\bar{C}(s_{n})$ from $\bar{C}(s_{n+1})$, we must first estimate the value of $\bar{C}(s_{\Omega})$ for some integer $\Omega$. In finding such an approximation the first point to note is that the value of $C_{i}(s_{\Omega})$ is predominantly determined by the value of $c_{i}(t)$ over the time-scale $0 \leq t \leq \frac{\tau}{\Omega}$: the value of $c_{i}(t)$ for $t > \frac{\tau}{\Omega}$ is relatively inconsequential. 

The second point to note is that by Equation (\ref{defn_St}), the cross-sectional area at time $\frac{\tau}{\Omega}$ is 
\begin{eqnarray*}
S_{ij} \big( \frac{\tau}{\Omega} \big) 
& = & S_{ij}(0) \big(2^{\frac{\Omega - 1}{\Omega}}  - 1 \big) +S_{ij}(\tau) \big(2 - 2^{\frac{\Omega - 1}{\Omega}} \big).
\end{eqnarray*}
\\
If $\Omega = 13$ (say), we have $S_{ij} \big( \frac{\tau}{\Omega} \big) \approx 0.9 S_{ij}(0) + 0.1 S_{ij}(\tau)$. We can estimate $C_{i}(s_{\Omega})$ by assuming that the cross-sectional area of each edge does not vary over time, but remains constant at $S_{ij} \big( \frac{\tau}{\Omega} \big)$. This enables us to apply Equation (41), and thereby obtain an estimate for $C_{i}( s_{\Omega})$. More specifically, for each node $i$ we make the initial guess that 
\begin{equation}
C_{i}( s_{\Omega})  =  \int_{0}^{\infty} c_{i}(t) e^{\frac{- \Omega \ln 2}{\tau}t} dt \quad \approx \quad \frac{ \tau  }{\Omega \ln 2} c_{i}(0),
\end{equation} 
(an approximation that would hold exactly if the concentration at the nodes remained constant). We can then employ the BiCGStab algorithm to home in on a more accurate solution to $\mathbf{M}( s_{\Omega}) \bar{C}( s_{\Omega}) = \bar{p}( s_{\Omega})$. Once we have obtained an estimate for $\bar{C}( s_{\Omega})$, we can  employ Equation (\ref{iterative_C}) to find $\bar{C}(s_{\Omega - 1}), \ldots , \bar{C}(s_{1})$. Also note that when the relative change in cross-sectional area is small, $b_{ij}$ is small compared to $a_{ij}$, so the elements in the vector $\mathbf{W}(s_{n+1})\bar{C}(s_{n})$ are small compared to the corresponding elements in $\mathbf{V}(s_{n})\bar{C}(s_{n})$ and $\bar{p}(s_{n})$. This means that when the relative change in cross-sectional area is small, any errors in our estimate for $\bar{C}( s_{\Omega})$ have little effect on the calculated values for $\bar{C}( s_{\Omega - 1})$.

\subsection{Resource distribution in a changing network} \label{resource for changing areas}
We now combine the preceding observations, and present an algorithm for calculating the concentration in each section of a network with changing cross-sectional areas after a given time $\tau$. More specifically, we suppose that for each edge in our network the mean velocity $u_{ij}$ and the dispersion coefficient $D_{ij}$ remain constant, while the cross-sectional areas $S_{ij}(t)$ vary smoothly from $S_{ij}(0)$ to $S_{ij}(\tau)$ in accordance with Equation (\ref{defn_St}). The first step in our algorithm is to pick an odd integer $\Omega$. This determines the scale of the errors in our final answer, and the ratio of the errors to the true values which will be of the order $\epsilon = 10^{-0.45(\Omega - 1)}$. As a rule of thumb setting $\Omega = 13$ and using standard double precision for the weights $\kappa_{n}$ will ensure that the our final answers are accurate to at least three significant digits. 

We let $s_{\Omega} = \frac{\Omega \ln 2}{\tau}$ and assume that the cross-sectional area of edge $ij$ is held constant at $S_{ij} \big( \frac{\tau}{\Omega} \big)$. These cross-sectional areas can be used to find a propagation matrix $\mathbf{M}(s_{\Omega})$, as described by Equation (25). Furthermore, the given initial condition can be used to calculate $\bar{p}(s_{\Omega})$, as described by Equations (42), (48) and (49).

Equation (41) tells us that $\mathbf{M}(s_{\Omega}) \bar{C}(s_{\Omega}) = \bar{p}(s_{\Omega})$, so we are now in a position to find $\bar{C}(s_{\Omega})$. More specifically, we can employ the BiCGStab algorithm, starting with the initial guess that for each $i$,
\begin{equation*}
C_{i}(s_{\Omega}) \approx \frac{\tau }{\Omega \ln 2 } c_{i}(0) = \frac{\tau }{\Omega \ln 2 } \frac{ \sum_{j} k_{ij}^{(1)} }{ \sum_{j} S_{ij}(0)}.
\end{equation*}
If the concentration at the nodes remained constant, this estimate would equal the exact solution. Once we have run the BiCGStab algorithm, we store the vector $\bar{C}(s_{\Omega})$, and use Equations (42), (48), (49), (\ref{defn_V}) and (\ref{defn_W}) to find $\mathbf{V}(s_{n})$, $\mathbf{W}(s_{n})$ and $\bar{p}(s_{n})$ where $n = \Omega -1$. 

These matrices are related to one another by Equation (\ref{iterative_C}), so once again we can employ the BiCGStab algorithm to find $\bar{C}(s_{n})$. As in the case of a network with constant cross-sectional areas, the matrices tend to be close to singular if any of the terms $\frac{l_{ij} \alpha_{ij}(s)}{2 D_{ij}}$ are large (greater than 10, say). This makes it computationally difficult to calculate $\bar{C}$, but this problem is easy to avoid as we can simply introduce additional nodes along the edge $ij$, thereby reducing the lengths $l_{ij}$. Having described the network in terms of nodes and sufficiently short edges, we start the BiCGStab algorithm with the initial guess that for each $i$,
\begin{equation*}
C_{i}(s_{n}) \approx \frac{n+1}{n} C_{i}(s_{n+1}).
\end{equation*}
This process is iterated until we have found $\bar{C}(s_{\Omega}), \ldots , \bar{C}(s_{1})$. Now, for each edge and each integer $n = \Omega - 1, \ldots , 1$, we can use Equations (\ref{defn_Qij_changing_S}) and (\ref{defn_Qji_changing_S}) to find the boundary conditions $X_{ij}(s_{n})$ and $X_{ji}(s_{n})$. As in the case of constant cross-sectional areas, the Laplace transform of resource that has reached a node is denoted by $Q_{ij}^{N}(x,s_{n})$, and this quantity is related to $X_{ij}(s_{n})$ and $X_{ji}(s_{n})$ by Equation (20). We can therefore use Equation (51) to find $Y_{ij}^{(n)}(s) = \mathcal{L} \big( y_{ij}^{(n)}(t) \big)$, the mean value of $Q_{ij}^{N}(x,s)$ in the $n$'th section of the edge $ij$.

Having found $Y_{ij}^{(n)}(s)$ for each section in the network, we can apply the Gaver-Stehfest algorithm to  obtain $y_{ij}^{(n)}(t)$. Finally, we note that the quantity of resource within each edge only changes because of advection, diffusion and delivery. Varying the cross-sectional area of an edge will affect the quantity of resource that enters that edge, but it will not directly affect the quantity or distribution of resource that remains in $ij$ without reaching nodes $i$ or $j$ over the time-scale $t$. For example, if an edge $ij$ is shrinking and the fluid within $ij$ is leaving that edge, it is assumed that the effect of this mass flow is entirely captured by calculating the appropriate velocity term $u_{ij}$. Given that this is so, for each edge $ij$ we can use Equations (54) and (56) to calculate a sequence of values for $e^{\lambda_{ij}^{m} t}$, $\mu_{ij}^{m}$ and $A^{m}$ until we reach an integer $\Omega'$ such that $e^{\lambda_{ij}^{\Omega'} t}$ satisfies Equation (58). This ensures that when we employ Equation (56) to find $z_{ij}^{(1)}(t), \ldots , z_{ij}^{(N_{ij})}(t)$  (the mean quantity of resource in $ij$ that has not reached a node), the errors are very small. Finally, we note that for each section of the network the mean quantity of resource per unit length
\begin{equation*}
k_{ij}^{(n)}(t) =  y_{ij}^{(n)}(t) + z_{ij}^{(n)}(t).
\end{equation*}

\subsection{Results}

We have presented two algorithms for calculating the concentration in a network as it changes over time, with resource subject to advection, diffusion and local delivery out of the network. In the first, cross-sectional areas and the medium-currents in each edge change in a stepwise manner, while in the second the cross-sectional areas and medium-currents in each edge vary continuously. The former algorithm is about twice as fast, and experimentation indicates that it is more numerically stable. 

By design, the total medium-current to pass through each edge in a given time step is the same whether the cross-sectional areas vary continuously or in a stepwise manner. Consequently, if the concentration at each node does not change dramatically over a given time step, the amount of resource to pass through each edge over the time step in question will be similar whichever of the two algorithms we apply. However, the two algorithms may give very different results for the concentration profile within a given edge, if there is a difference in the initial concentration at the nodes at either end. 


\begin{figure}[h!]
\begin{center}
\includegraphics[width=8.5cm]{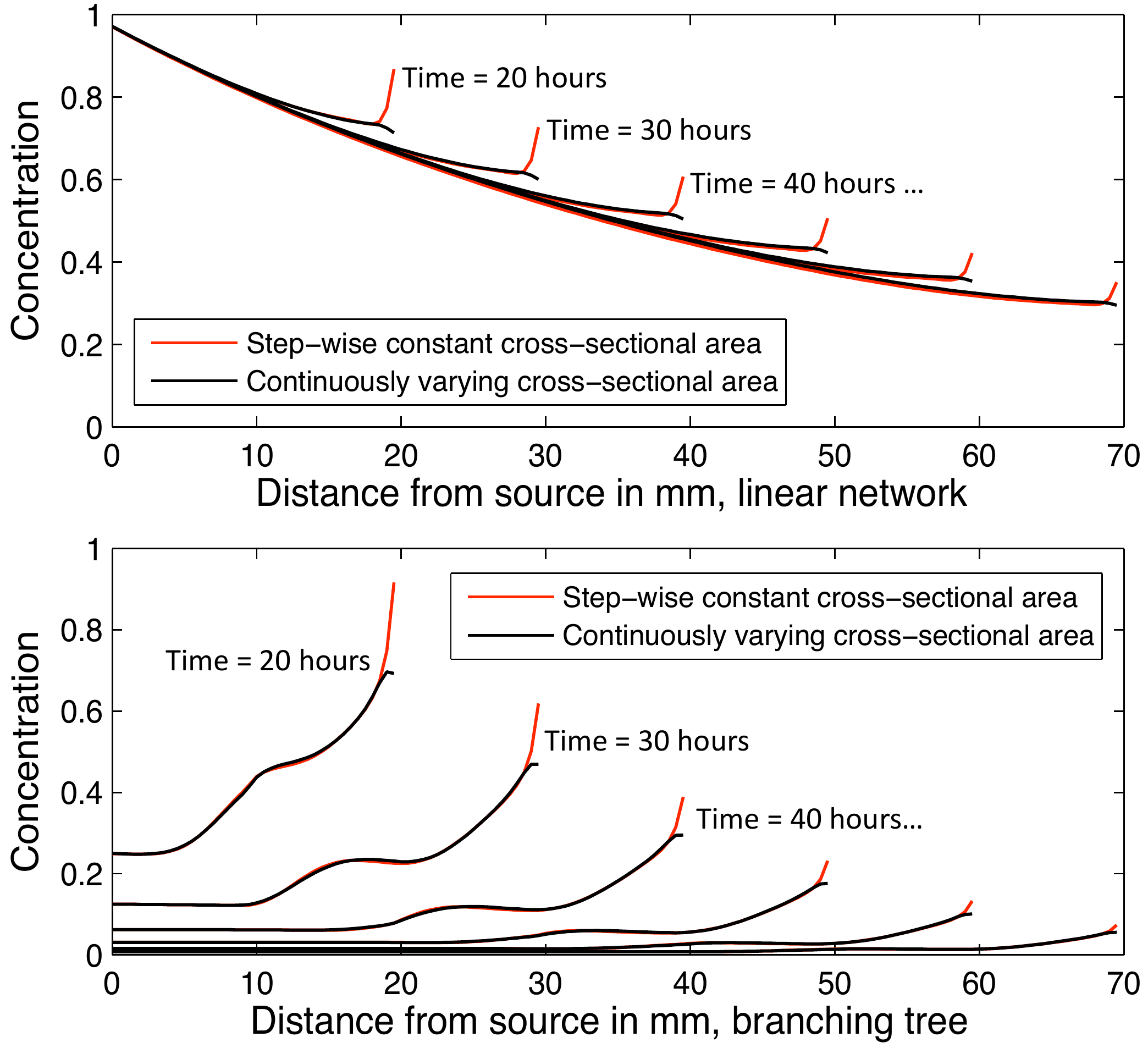}
\caption{\label{conc_linear_and_branching_tree}  \textbf{Concentration in a linear network and branching tree.} In each case resource is added at the source at a constant rate, the local delivery rate throughout each network is $0.02 \textrm{hour}^{-1}$, and the diffusion coefficient is $D_{m} = 6.7 \times 10^{-4} \textrm{mm}^{2}\textrm{s}^{-1}$. Every hour, both networks grow $1 \textrm{mm}$ further from the source, and the branching tree bifurcates every $10 \textrm{mm}$. The cross-sectional area of each new edge either varies in a stepwise manner from 0 to $6 \mu\textrm{m}$ (red lines), or the cross-sectional area varies continuously from 0 to $6 \mu\textrm{m}$ (black lines). Note that the same amount of resource is contained within the linear network and the branching tree, but the branching tree has a greater volume. After 70 hours the mean concentration in the linear network is over 18 times greater than the mean concentration in the branching tree, but the concentration near the tips is only about 5 times as great.}
\end{center}
\end{figure}

\newpage
\nocite{*}
\bibliographystyle{abbrv}
\bibliography{ADC_network}

\end{document}